\documentclass[12pt, draftclsnofoot, onecolumn]{IEEEtran}
\usepackage{cite}
\usepackage{amssymb,amsfonts}
\usepackage{algorithmic}
\usepackage{graphicx}
\usepackage{textcomp}
\usepackage{amsmath}
\usepackage{amssymb}
\usepackage{latexsym}
\usepackage{CJK}
\usepackage{booktabs}
\usepackage{array}
\usepackage{multirow}
\usepackage{graphicx}
\usepackage{epstopdf}
\usepackage{caption}

\usepackage{algorithm}
\usepackage{algorithmic}
\usepackage{tabularx}
\usepackage{stfloats}
\usepackage{xcolor}
\usepackage{booktabs}

\usepackage{makecell}
\usepackage{threeparttable}
\usepackage{bm}
\usepackage{subfig}
\usepackage{float}

\usepackage[colorlinks=false,linkcolor=black]{hyperref} 

\usepackage{amsbsy}

\usepackage{bbding}

\allowdisplaybreaks[4]
\usepackage{setspace}

\ifCLASSINFOpdf
\else
\fi

\begin{document}
%
\title{Cooperative Beamforming for RIS-Aided Cell-Free Massive MIMO Networks}


\author{
       Xinying Ma, Deyou Zhang, Ming Xiao, \IEEEmembership{Senior Member, IEEE}, \\ Chongwen Huang, \IEEEmembership{Member, IEEE}, and Zhi Chen, \IEEEmembership{Senior Member, IEEE}  \vspace{-2.5em}
\thanks{
X. Ma and Z. Chen are with the National Key Laboratory of Science and Technology on Communications, University of Electronic Science and Technology of China, Chengdu, China (e-mails: xymarger@126.com, chenzhi@uestc.edu.cn).

D. Zhang and M. Xiao are with the School of Electrical Engineering and Computer Science, KTH Royal Institute of Technology, Stockholm, Sweden (e-mails: deyou@kth.se, mingx@kth.se).

C. Huang is with the College of Information Science and Electronic Engineering, Zhejiang University, Hangzhou, China (e-mail: chongwenhuang@zju.edu.cn).
}
}

\maketitle

\begin{abstract}
The combination of cell-free massive multiple-input multiple-output (CF-mMIMO) and reconfigurable intelligent surface (RIS) is envisioned as a promising paradigm to improve network capacity and enhance coverage capability. However, to reap full benefits of RIS-aided CF-mMIMO, the main challenge is to efficiently design cooperative beamforming (CBF) at base stations (BSs), RISs, and users. Firstly, we investigate the fractional programing to convert the weighted sum-rate (WSR) maximization problem into a tractable optimization problem. Then, the alternating optimization framework is employed to decompose the transformed problem into a sequence of subproblems, i.e., hybrid BF (HBF) at BSs, passive BF at RISs, and combining at users. In particular, the alternating direction method of multipliers algorithm is utilized to solve the HBF subproblem at BSs. Concretely, the analog BF design with unit-modulus constraints is solved by the manifold optimization (MO) while we obtain a closed-form solution to the digital BF design that is essentially a convex least-square problem. Additionally, the passive BF at RISs and the analog combining at users are designed by primal-dual subgradient and MO methods. Moreover, considering heavy communication costs in conventional CF-mMIMO systems, we propose a partially-connected CF-mMIMO (P-CF-mMIMO) framework to decrease the number of connections among BSs and users. To better compromise WSR performance and network costs, we formulate the BS selection problem in the P-CF-mMIMO system as a binary integer quadratic programming (BIQP) problem, and develop a relaxed linear approximation algorithm to handle this BIQP problem. Finally, numerical results demonstrate superiorities of our proposed algorithms over baseline counterparts.
\end{abstract}

\begin{IEEEkeywords}
Cell-free massive multiple-input multiple-output (CF-mMIMO), reconfigurable intelligent surface (RIS), cooperative beamforming (CBF), base station selection, integer programming.
\end{IEEEkeywords}


\section{Introduction}
The deployment of ultra-dense network (UDN) has been considered as an indispensable technology to meet the requirements of massive connectivity and seamless coverage for sixth generation (6G) wireless networks \cite{introduction01}. Intriguingly, the basic idea of UDN can be well achieved by an emerging embodiment, namely cell-free massive multiple-input multiple-output (CF-mMIMO)\cite{introduction02,introduction02a,introduction02b}, which benefits from the integration of distributed networks and massive MIMO.

With regard to CF-mMIMO networks, a considerable number of distributed base stations (BSs) coherently serve a small number of users on the same time-frequency resources and all the BSs are connected to a central processing unit (CPU) via backhaul wireless links \cite{introduction03}. There is no channel state information (CSI) exchange among these BSs, and thus the CPU acts an important role in coordination and computational assistance for BSs. Under such a setup, CF-mMIMO entails some distinctive advantages as follows: i) thanks to a large number of BSs, CF-mMIMO provides much higher coverage probability and stronger macro-diversity than conventional small-cell networks \cite{introduction04}; ii) since CF-mMIMO networks possess no cell boundaries, the inter-cell interference (ICI) can be eliminated by jointly cooperating the distributed BSs \cite{introduction02}; iii) favorable propagation is another potential characteristic existing in CF-mMIMO networks that is able to alleviate inter-user interferences (IUIs) \cite{introduction05}. Nevertheless, in addition to these advantages, CF-mMIMO also faces some challenges in practice.

On the one hand, high frequency bands adopted by CF-mMIMO in 6G wireless networks are more vulnerable to blockage, and thus much denser deployment of BSs is demanded to compensate for the coverage hole and the serious propagation attenuation, e.g., millimeter wave (mmWave) and terahertz bands \cite{introduction06,introduction06a}. However, more distributed BSs definitely result in much higher network overhead. To balance the network capacity and energy consumption, the reconfigurable intelligent surface (RIS) is regarded as a promising paradigm to assist mmWave CF-mMIMO networks in an energy-efficient way \cite{introduction07,introduction08,introduction09}. Specifically, RIS consists of a large number of passive reflecting elements and is capable of manipulating the propagation direction of impinging electromagnetic waves by adjusting the amplitude and phase shift without active radio frequency (RF) chains \cite{introduction09a,introduction09b,introduction09c}. With the assistance of RISs, the wireless environment becomes controllable and reconfigurable, and CF-mMIMO networks can enhance the signal coverage capability with low energy consumption \cite{introduction09d}.
Therefore, the combination of RIS and CF-mMIMO is cast as a potential approach to improve the network capability. To fully reap the benefits, the cooperative beamforming (CBF) design (e.g., at BSs, RISs and users) is a crucial performance indicator for RIS-aided mmWave CF-mMIMO networks. In \cite{introduction10}, authors first estimate the CSI by sending uplink pilots, and then design the conjugate BF for downlink data transmission with estimated CSI. To maximize the energy efficiency, the work in \cite{introduction11} proposes a joint design of transmit beamformers at BSs and reflecting coefficients at RISs in the case of limited backhaul capacity. In \cite{introduction12}, a fully decentralized BF scheme is originally developed, which greatly decreases the backhaul signaling compared to centralized approaches. However, these previous works all consider single-antenna users and low frequency bands that are unable to meet the requirements of 6G wireless networks. Although the multi-antenna users are considered in the RIS-aided CF-mMIMO network \cite{introduction13}, this work neglects the combining design at users and lacks the consideration for hybrid BF (HBF) at BSs. Hence, it is imperative to explore the efficient CBF design for RIS-aided mmWave CF-mMIMO networks.

On the other hand, it is impractical for conventional mmWave CF-mMIMO networks that all users are served by all BSs. This is because fully-connected mode endures serious communication costs, including but not limited to backhaul signaling overhead, energy consumption and intensive computation \cite{introduction14}. Fortunately, one of the most promising solutions is to consider the partially-connected CF-mMIMO (P-CF-mMIMO), also called user-centric approach, where each user is only served by a subset of BSs \cite{introduction15,introduction16,introduction17}. By properly cutting down the number of connected links among users and BSs, the P-CF-mMIMO network is capable of suppressing communication costs with negligible performance degradation compared to the conventional CF-mMIMO network. For instance, considering the P-CF-mMIMO network, the work in \cite{introduction16} considers that each BS only selects a part of users that own stronger channel gains (e.g., larger norms). This simple selection criterion brings obvious performance penalty in the case of perfect CSI and erodes fairness for users with weak channel conditions. In \cite{introduction17}, a newly scalable CF-mMIMO system is proposed by exploiting the dynamic cooperation cluster, but neglects the diagonal matrix optimization in terms of BS selection. In addition, the work in \cite{introduction18} proposes a structured massive access framework for CF-mMIMO systems, where each user selects its desired BSs by a competitive mechanism. This heuristic BS selection strategy is time-consuming and cannot guarantee optimal network performance. It is worth noting that these existing works are lack of discovering intrinsic BS selection peculiarities  \cite{introduction16,introduction17,introduction18,introduction18a}. More precisely, the BS selection problem can be treated as a convex integer optimization problem, which can be efficiently settled from the perspective of integer programming.

Motivated by above discussion, we first focus on the joint BF design of passive BF at RISs and active BF at transceivers by maximizing the weighted sum-rate (WSR) for the RIS-aided mmWave CF-mMIMO system. Then, the second implementation is to address the P-CF-mMIMO network and solve the intractable BS selection problem in the way of integer programming. The main contributions can be listed as follows.
\begin{itemize}
  \item We propose a general RIS-aided mmWave CF-mMIMO system, where multiple BSs with multiple antennas serve multiple users with multiple antennas assisted by multiple RISs. Then we transform the non-convex and multiple-ratio WSR maximization problem into a sequence of tractable subproblems by leveraging the fractional programming (FP).
  \item To achieve the HBF design at BSs, an alternating direction method of multipliers (ADMM) algorithm is utilized to optimize the augmented Lagrangian objective with non-convex constraints. Specifically, we use manifold optimization (MO) to design the analog BF (ABF) with unit-modulus constraints and derive a closed-form solution for digital BF (DBF) design.
  \item We formulate the passive BF design at RISs as a convex quadratically constrained quadratic program (QCQP) problem and directly employ the primal-dual subgradient (PDS) algorithm to optimize reflecting coefficients of RISs. Since the combining design at users is recast as a non-convex QCQP problem with unit-modulus constraints, the MO algorithm is utilized to obtain a high-quality solution.
  \item In addition, we originally model the BS selection problem as a binary integer quadratic programming (BIQP) problem in the RIS-aided P-CF-mMIMO network, and develop a relaxed linear approximation (RLA) algorithm to solve this formulated BIQP problem by substituting the quadratic term with new variables and linear constraints.
  \item Simulation results reveal that RISs can enhance the WSR performance of both CF-mMIMO and P-CF-mMIMO networks compared to the case without RISs and the case with random phase of RISs. Remarkably, the RLA empowered P-CF-mMIMO network basically approaches the WSR performance of the conventional CF-mMIMO network and meanwhile decreases communication costs.
\end{itemize}

The remainder of this paper is organized as follows. The RIS-enabled CF-mMIMO system model and problem formulation are presented in Section II. In Section III, we explore the CBF design. Section IV discusses the RLA based BS selection for the P-CF-mMIMO system. Finally, simulation results and conclusion are presented in Section V and Section VI, respectively.

\emph{Notations:} ${{{\bf{A}}^H}}$, ${{\bf{A}}^ * }$, ${{{\bf{A}}^T}}$, ${{{\bf{A}}^{ - 1}}}$, ${{{\bf{A}}^\dag }}$ and ${\rm{rank}}\left( {\bf{A}} \right)$ are conjugate transpose, conjugate, transpose, inverse, pseudo-inverse and the rank of ${\bf{A}}$, respectively. ${{\left\| {\bf{A}} \right\|_F}}$ is the Frobenius norm. ${\rm{diag}} \left( {\bf{a}} \right)$ is a diagonal matrix with elements of ${\bf{a}}$ on its diagonal. ${{\rm{tr}}} \left( {\bf{A}} \right)$  is the trace of matrix ${\bf{A}}$.  ${\rm{vec}} \left( {\bf{A}} \right)$ is the column-ordered vectorization. ${{\bf{1}}_N}$ and ${{\bf{0}}_N}$ denote $N$-dimensional all-ones and all-zeros vector. $\Re(\cdot)$ and $\Im(\cdot)$ denote the real and imaginary part of its argument. $\otimes$ and $\circ$ denote the Kronecker and Hadamard products.  ${\mathbb E} \left[  \cdot  \right]$ denotes the expectation, and ${{\cal O}}\left( \cdot \right)$ indicates the number of complex multiplications.


\section{System Model of RIS-Aided CF-mMIMO Networks}
In this section, we introduce the RIS-aided CF-mMIMO system model as well as the formulation of the WSR maximization problem.

\subsection{System Model}

\begin{figure}[!t]
\centering
\includegraphics[width=7 cm]{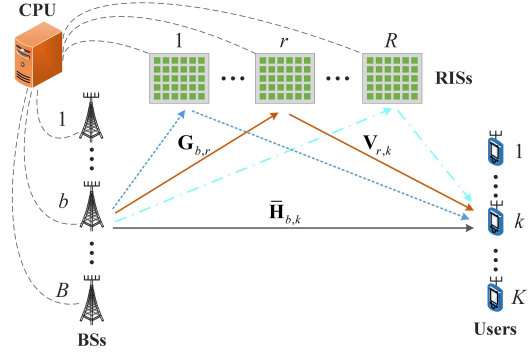}
\caption{Illustration of RIS-aided CF-mMIMO network.}\label{CFSystem} \vspace{-1.5em}
\end{figure}

We consider a downlink RIS-aided CF-mMIMO network, as shown in Fig. \ref{CFSystem}, where a set of BSs ${\cal B} = \{ 1,2, \cdots ,B\}$  serve a group of multi-antenna users ${\cal K} = \{ 1,2, \cdots ,K\}$  via the aid of a set of RISs ${\cal R} = \{ 1,2, \cdots ,R\}$. Each BS with $N_{\text{RF}}$ RF chains, each user with one RF chain and each RIS without RF chain are equipped with $N_t$, $N_r$ and $M$  array elements, respectively. Besides,  all BSs and RISs are connected to the CPU to implement signal processing. Assume that BSs employ the HBF architecture and each BS transmits $K$ data streams to the users. In the downlink mode, the transmitted signal from $b$th BS can be expressed as
\begin{align}\label{CFsystem01}
{{\bf{x}}_b} = {{\bf{F}}_{{\rm{RF,}}b}}{{\bf{F}}_{{\rm{BB,}}b}}{\bf{s}} = \sum\limits_{k = 1}^K {{{\bf{F}}_{{\rm{RF,}}b}}{{\bf{f}}_{{\rm{BB,}}b,k}}{s_k}},
\end{align}
where ${\bf{s}} \in {[{s_1},{s_1}, \cdots ,{s_K}]^T} \in {{\mathbb C}^{K \times 1}}$ indicates the symbol vector that satisfies ${\mathbb E}[{\bf{s}}{{\bf{s}}^H}] = {{\bf{I}}_K}$.   ${{\bf{F}}_{{\rm{RF,}}b}} \in {{\mathbb C}^{{N_t} \times {N_{{\rm{RF}}}}}}$ and ${{\bf{F}}_{{\rm{BB,}}b}} = \left[ {{{\bf{f}}_{{\rm{BB,}}b,1}}, \cdots ,{{\bf{f}}_{{\rm{BB,}}b,K}}} \right] \in {{\mathbb C}^{{N_{{\rm{RF}}}} \times K}}$ denotes ABF and DBF matrices at $b$th BS, respectively. The power constraint for $b$th BS is denoted by  $||{\bf{F}}_{{\rm{RF}},b}{\bf{F}}_{{\rm{BB}},b}||_F^2 \le {P_b}$ $(\forall b \in {\cal B})$, where $P_b$  is the maximum transmit power.

Due to the deployment of RISs, the entire channel model between $b$th BS and $k$th user is composed of BS-user link ${{\bf{\bar H}}_{b,k}} \in {{\mathbb C}^{{N_r} \times {N_t}}}$, BS-RIS link ${{\bf{G}}_{b,r}} \in {{\mathbb C}^{M \times {N_t}}}$  and RIS-user link ${{\bf{V}}_{r,k}} \in {{\mathbb C}^{{N_r} \times M}}$. Hence, the equivalent channel from $b$th BS to $k$th user can be expressed as
\begin{align}\label{CFsystem02}
{{\bf{H}}_{b,k}} = {{\bf{\bar H}}_{b,k}} + \sum\limits_{r = 1}^R {{{\bf{V}}_{r,k}}{{\bf{\Theta }}_r}{{\bf{G}}_{b,r}}},
\end{align}
where ${{\bf{\Theta }}_r} = {{\rm{diag}}}({\theta _{r,1}}, \cdots ,{\theta _{r,M}}) \in {{\mathbb C}^{M \times M}}$ denotes the phase shift matrix of $r$th RIS and each diagonal entry can be further defined as ${\theta _{r,m}} = {u_{r,m}}{e^{j{\phi _{r,m}}}}, r \in {\cal R},m \in {\cal M}$. In practice, the reflecting amplitude ${u_{r,m}}$  and phase shift ${\phi _{r,m}}$ can be controlled separately, where $|{u_{r,m}}| \le 1$. Considering the sparse nature of mmWave channel, the typical Saleh-Valenzuela channel model with limited scattering paths is adopted \cite{introduction19}. Thus, ${{\bf{\bar H}}_{b,k}}$, ${{\bf{G}}_{b,r}}$ and ${{\bf{V}}_{r,k}}$  can be represented as
\begin{align}\label{CFsystem03}
{\bf{H}}_{S} = \sqrt {\frac{{{N_1}{N_2}}}{{L}}} \sum\limits_{l = 1}^L {{\beta _l}{{\bf{a}}_2}\left( {\omega _l^2,\varsigma _l^2} \right){\bf{a}}_1^H\left( {\omega _l^1,\varsigma _l^1} \right)},
\end{align}
where ${\bf{H}}_{S} \in \left\{ {{\bf{\bar H}}_{b,k}}, {{\bf{G}}_{b,r}}, {{\bf{V}}_{r,k}} \right\}$. $L$ is the number of paths that contains one LoS path and $(L-1)$  NLoS paths, and ${\beta _l}$ denotes the complex path gain of $l$th path. $N_1$  and  $N_2$ respectively denote the numbers of array elements at transmitter and receiver. $\omega _l^1(\varsigma _l^1)$  and $\omega _l^2(\varsigma _l^2)$  denote the associated azimuth (elevation) angles of arrival (AoAs) and angles of departure (AoDs) of $l$th path, respectively. In addition, the uniform planar array (UPA) structure is considered for BSs, RISs and users. The array response for the UPA with ${N_x}{N_y}-$elements can be given by
\begin{align}\label{CFsystem04}
{\bf{a}}\left( {\omega ,\varsigma } \right) = \frac{1}{{\sqrt {{N_x}{N_y}} }}{[\cdots ,{e^{j\frac{{2\pi }}{{{\lambda _s}}}{d_s}({n_x}\sin (\omega )\sin (\varsigma ) + {n_y}\cos (\varsigma ))}}, \cdots ]^T},
\end{align}
where $0 \le {n_x} \le {N_x} - 1$  and $0 \le {n_y} \le {N_y} - 1$. ${\lambda _s}$ and ${d_s}$ represent the signal wavelength and the antenna spacing, respectively.
By combining (\ref{CFsystem01}) and (\ref{CFsystem02}), the downlink received signal from $b$th BS to $k$th user can be expressed as
\begin{align}\label{CFsystem05}
{{\bf{y}}_{b,k}} = \sum\limits_{j = 1}^K {{{\bf{H}}_{b,k}}{{\bf{F}}_{{\rm{RF,}}b}}{{\bf{f}}_{{\rm{BB,}}b,j}}{s_j}}.
\end{align}

Since $B$ BSs send data streams to $K$ users simultaneously, the received signal at $k$th user is the superposition of the signals transmitted by $B$ BSs. Hence, the received signal at $k$th user from all the BSs can be expressed as
\begin{align}\label{CFsystem06}
{{\bf{y}}_k} = \sum\limits_{b = 1}^B {{{\bf{y}}_{b,k}}}  + {{\bf{n}}_k}
= \sum\limits_{b = 1}^B {\sum\limits_{j = 1}^K {{{\bf{H}}_{b,k}}{{\bf{F}}_{{\rm{RF,}}b}}{{\bf{f}}_{{\rm{BB,}}b,j}}{s_j}} }  + {{\bf{n}}_k},
\end{align}
where ${{\bf{n}}_k}$ denotes the noise at $k$th user following Gaussian distribution of ${\cal C}{\cal N}\left( {{\bf{0}},\sigma _c^2{{\bf{I}}_{{N_r}}}} \right)$ that corrupts the received signal. After being processed by a combiner ${{\bf{w}}_k} \in {{\mathbb C}^{{N_r} \times 1}}$, the received signal of $k$th user can be further written as
\begin{align}\label{CFsystem07}
{{\tilde y}_k} & = \underbrace {\sum\limits_{b = 1}^B {{\bf{w}}_k^H\left( {{{{\bf{\bar H}}}_{b,k}} + {{\bf{V}}_k}{\bf{\Phi }}{{\bf{G}}_b}} \right){{\bf{F}}_{{\rm{RF,}}b}}{{\bf{f}}_{{\rm{BB,}}b,k}}{s_k}} }_{\text{Useful signal}} \notag\\
&  + \underbrace {\sum\limits_{b = 1}^B {\sum\limits_{j = 1,j \ne k}^K {{\bf{w}}_k^H\left( {{{{\bf{\bar H}}}_{b,k}} + {{\bf{V}}_k}{\bf{\Phi }}{{\bf{G}}_b}} \right){{\bf{F}}_{{\rm{RF,}}b}}{{\bf{f}}_{{\rm{BB,}}b,j}}{s_j}} } }_{\text{Interference}} + \underbrace {{\bf{w}}_k^H{{\bf{n}}_k}}_{\text{Noise}}.
\end{align}
where ${{\bf{V}}_k} = [{{\bf{V}}_{1,k}}, \cdots ,{{\bf{V}}_{R,k}}] \in {{\mathbb C}^{{N_r} \times RM}}$, ${{\bf{G}}_b} = {[{\bf{G}}_{b,1}^T, \cdots ,{\bf{G}}_{b,R}^T]^T} \in {{\mathbb C}^{RM \times {N_t}}}$ and ${\bf{\Phi }} = {\rm{diag}}\left( {{{\bf{\Theta }}_1}, \cdots ,{{\bf{\Theta }}_R}} \right) \in {{\mathbb C}^{RM \times RM}}$, respectively.

\subsection{Problem Formulation}
To maximize the WSR of the RIS-aided CF-mMIMO system, the signal-to-interference-plus-noise ratio (SINR) needs to be calculated for each user. Based on (\ref{CFsystem07}), the SINR of $k$th user can be expressed as
\begin{align}\label{CFsystem09}
{\gamma _k} = \frac{{{{\left| {\sum\limits_{b = 1}^B {{\bf{w}}_k^H{{\bf{H}}_{b,k}}{{\bf{F}}_{{\rm{RF,}}b}}{{\bf{f}}_{{\rm{BB,}}b,k}}} } \right|}^2}}}{{\sum\limits_{j = 1,j \ne k}^K {{{\left| {\sum\limits_{b = 1}^B {{\bf{w}}_k^H{{\bf{H}}_{b,k}}{{\bf{F}}_{{\rm{RF,}}b}}{{\bf{f}}_{{\rm{BB,}}b,j}}} } \right|}^2}}  + {\sigma ^2}}},
\end{align}
where ${\sigma ^2} = N_r{\sigma _c^2}$ denotes the effective noise variance.

Subsequently, the WSR of the RIS-aided CF-mMIMO system can be written as
\begin{align}\label{CFsystem10}
R_{sum} = \sum\limits_{k = 1}^K {{\omega _k}\log \left( {1 + {\gamma _k}} \right)},
\end{align}
where ${\omega _k} > 0$ $(\forall k \in {\cal K})$ denotes the WSR weight for $k$th user with $\sum\nolimits_{k \in {\cal K}} {{\omega _k}}  = 1$.

Finally, the WSR maximization problem can be formulated as
\begin{subequations}\label{CFsystem11}
\begin{align}
& \mathop {\max}\limits_{{\bf{F}},{\bf{\Phi}},{\bf{w}}} \; {R_{sum}} \label{CFsystem11main} \\
& \;\; {\rm{s.t.}} \;\; \left\| {{{\bf{F}}_{{\rm{RF,}}b}}{{\bf{F}}_{{\rm{BB,}}b}}} \right\|_F^2 \le {P_b},\forall b \in {\cal B},\label{CFsystem11a}\\
& \;\;\;\;\;\;\;\;\;\; {\left| {{\bf{F}}_b^{{\rm{RF}}}\left( {{i_1},{j_1}} \right)} \right|^2} = 1,\forall {i_1},{j_1},b \in {\cal B},\label{CFsystem11b}\\
& \;\;\;\;\;\;\;\;\;\; {\left| {{{\bf{\Theta }}_r}\left( {{i_2},{i_2}} \right)} \right|^2} \le 1,\forall {i_2},r \in {\cal R},\label{CFsystem11c}\\
& \;\;\;\;\;\;\;\;\;\; {\left| {{{\bf{w}}_k}\left( {{i_3}} \right)} \right|^2} = 1,\forall {i_3},k \in {\cal K}, \label{CFsystem11d}
\end{align}
\end{subequations}
where ${\bf{F}} = \{ {{\bf{F}}_b}| \; \forall b \in {\cal B}\}$, ${{\bf{F}}_b} = {{\bf{F}}_{{\rm{RF,}}b}}{{\bf{F}}_{{\rm{BB,}}b}}$ and ${\bf{w}} = {[{\bf{w}}_1^T,{\bf{w}}_2^T, \cdots ,{\bf{w}}_K^T]^T}$, respectively.

\section{Cooperative Beamforming for RIS-Aided CF-mMIMO Networks}
The purpose of this section is to realize the CBF design for the RIS-aided CF-mMIMO network. To convert problem (\ref{CFsystem11}) into a tractable problem, we use the Lagrangian dual transform and FP transformation \cite{System01} to deal with the sum-logarithm term, which can be given by
\begin{align}\label{CFsystem12}
{f_{\rm{D}}}\left( {{\bf{F}},{\bf{\Phi }},{\bf{w}},{\bm{\lambda }}} \right) =  \sum\limits_{k = 1}^K {\left( {{\omega _k}\log \left( {1 + {\lambda _k}} \right)} - {{\omega _k}{\lambda _k}} + {\frac{{{\mu _k}{{\left| {\sum\limits_{b = 1}^B {{\bf{w}}_k^H{{\bf{H}}_{b,k}}{{\bf{F}}_{{\rm{RF,}}b}}{{\bf{f}}_{{\rm{BB,}}b,k}}} } \right|}^2}}}{{\sum\limits_{j = 1}^K {{{\left| {\sum\limits_{b = 1}^B {{\bf{w}}_k^H{{\bf{H}}_{b,k}}{{\bf{F}}_{{\rm{RF,}}b}}{{\bf{f}}_{{\rm{BB,}}b,j}}} } \right|}^2}}  + {\sigma ^2}}}} \right)},
\end{align}
where ${\mu _k} = {\omega _k}\left( {1 + {\lambda _k}} \right)$ and ${\bm{\lambda }} = \{ {\lambda _k}| \; \forall k \in {\cal K}\}$ refers to a collection of auxiliary variables.

By using the quadratic transform on the fractional term, we further recast ${f_{\rm{D}}}\left( {{\bf{F}},{\bf{\Phi }},{\bf{w}},{\bm{\lambda }}} \right)$ as
\begin{align}\label{CFsystem13}
{f_{\rm{Q}}}\left( {{\bf{F}},{\bf{\Phi }},{\bf{w}},{\bm{\lambda }},{\bm{\xi }}} \right) & =\sum\limits_{k = 1}^K {{\omega _k}\log \left( {1 + {\lambda _k}} \right)}  + \Re \left( {\sum\limits_{k = 1}^K {\sum\limits_{b = 1}^B {2\sqrt {{\mu _k}} \xi _k^H{\bf{w}}_k^H{{\bf{H}}_{b,k}}{{\bf{F}}_{{\rm{RF,}}b}}{{\bf{f}}_{{\rm{BB,}}b,k}}} } } \right) \notag \\
& - \sum\limits_{k = 1}^K {{\omega _k}{\lambda _k}} - \sum\limits_{k = 1}^K {{{\left| {{\xi _k}} \right|}^2}{\sigma ^2}}  - \sum\limits_{k = 1}^K {\sum\limits_{j = 1}^K {{{\left| {{\xi _k}} \right|}^2}{{\left| {\sum\limits_{b = 1}^B {{\bf{w}}_k^H{{\bf{H}}_{b,k}}{{\bf{F}}_{{\rm{RF,}}b}}{{\bf{f}}_{{\rm{BB,}}b,j}}} } \right|}^2}}},
\end{align}
where ${\bm{\xi }} = \{ {\xi _k}|\; \forall k \in {\cal K}\}$ denotes a set of auxiliary variables.

Subsequently, we define $f\left( {{\bf{F}},{\bf{\Phi }},{\bf{w}},{\bm{\lambda }},{\bm{\xi }}} \right) =  - {f_{\rm{Q}}}\left( {{\bf{F}},{\bf{\Phi }},{\bf{w}},{\bm{\lambda }},{\bm{\xi }}} \right)$, and problem (\ref{CFsystem11}) can be equivalently rewritten as
\begin{subequations}\label{CFsystem14}
\begin{align}
& \mathop {{\rm{min}}}\limits_{{\bf{F}},{\bf{\Phi }},{\bf{w}}} \; f\left( {{\bf{F}},{\bf{\Phi}},{\bf{w}},{\bm{\lambda}},{\bm{\xi}}} \right) \label{CFsystem14main} \\
& \;\; {\rm{s.t.}} \;\; {\text{(\ref{CFsystem11a}), (\ref{CFsystem11b}), (\ref{CFsystem11c}), (\ref{CFsystem11d})}}.
\end{align}
\end{subequations}

It can be observed that constraints (\ref{CFsystem11a})-(\ref{CFsystem11d}) involve multiple variables and include non-convex sets, e.g., (\ref{CFsystem11b}), (\ref{CFsystem11d}). To tackle this issue, the alternating optimization (AO) is widely treated as an efficient approach \cite{System01a}. More specifically, we optimize BF variables $\left( {{\bf{F}},{\bf{\Phi }},{\bf{w}}} \right)$ and auxiliary variables $\left( {{\bm{\lambda }},{\bm{\xi }}} \right)$ in an iterative manner until the objective function converges.

\subsection{Fix $\left( {{\bf{F}},{\bf{\Phi }},{\bf{w}}} \right)$ and Solve \protect\boldmath $\left( {{\bf{\lambda }},{\bf{\xi}}} \right)$}
According to the FP transform process in \cite{System01}, the optimal solutions to $\left( {{\bm{\lambda }},{\bm{\xi }}} \right)$ can be respectively calculated by solving ${{\partial {f_{\rm{D}}}\left( {{\bf{F}},{\bf{\Phi }},{\bf{w}},{\bm{\lambda }},{\bm{\xi }}} \right)}}/{{\partial {\lambda _k}}} = 0$ and ${{\partial {f_Q}\left( {{\bf{F}},{\bf{\Phi }},{\bf{w}},{\bm{\lambda }},{\bm{\xi }}} \right)}}/{{\partial {\xi _k}}} = 0$, $\forall k \in {\cal K}$. Then, the optimal variables $({{\bm{\lambda }}^{\star}},{{\bm{\xi }}^{\star}})$ can be respectively given by
\begin{align}\label{SecIIIHB01}
\lambda _k^{\star} = {\gamma _k}, \;\;\;\;\;\;\;\; \xi _k^{\star} = \frac{{\sqrt {{\mu _k}} \left( {\sum\limits_{b = 1}^B {{\bf{w}}_k^H{{\bf{H}}_{b,k}}{{\bf{F}}_{{\rm{RF,}}b}}{{\bf{f}}_{{\rm{BB,}}b,k}}} } \right)}}{{\sum\limits_{j = 1}^K {\left( {{{\left| {\sum\limits_{b = 1}^B {{\bf{w}}_k^H{{\bf{H}}_{b,k}}{{\bf{F}}_{{\rm{RF,}}b}}{{\bf{f}}_{{\rm{BB,}}b,j}}} } \right|}^2}} \right)}  + {\sigma ^2}}}, \;\; \forall k \in {\cal K}.
\end{align}

With the optimized  $({{\bm{\lambda }}^{\star}},{{\bm{\xi }}^{\star}})$, the CBF optimization problem is divided into three subproblems that can be settled alternately.

\subsection{Fix \protect\boldmath$\left( {{\bf{\Phi }},{\bf{w}},{\bf{\lambda }},{\bf{\xi }}} \right)$  and Solve ${\bf{F}}$}
To begin with, the considered CF-mMIMO network belongs to a centralized mode, which implies that all the related information can be exchanged and acquired by the CPU. Therefore, the HBF design for $B$ BSs can be executed in parallel. For $b$th $(\forall b \in {\cal B})$ BS, the reformulated WSR optimization problem can be given by
\begin{subequations}\label{SecIIIHB04}
\begin{align}
& \mathop {{\rm{min}}}\limits_{{{\bf{F}}_{{\rm{RF,}}b}},{{\bf{F}}_{{\rm{BB,}}b}}} \; {f_1}\left( {{{\bf{F}}_{{\rm{RF,}}b}},{{\bf{F}}_{{\rm{BB,}}b}}} \right) \label{SecIIIHB04main} \\
& \;\;\;\;\; {\rm{s.t.}} \;\; {\text{(\ref{CFsystem11a}),\;(\ref{CFsystem11b}),}}
\end{align}
\end{subequations}
where the new objective function ${f_1}\left( {{\bf{F}}_{{\rm{RF}},b},{\bf{F}}_{{\rm{BB}},b}} \right)$ is given by
\begin{align}\label{SecIIIHB06}
{f_1}\left( {{\bf{F}}_{{\rm{RF}},b},{\bf{F}}_{{\rm{BB}},b}} \right) & = \sum\limits_{k = 1}^K {\sum\limits_{j = 1}^K {{{\left| {{\xi _k}} \right|}^2}{{\left| {{\bf{w}}_k^H{{\bf{H}}_{b,k}}{{\bf{F}}_{{\rm{RF,}}b}}{{\bf{f}}_{{\rm{BB,}}b,j}} + {C_{b,k,j}}} \right|}^2}} } \notag \\
& -2\Re \left( {\sum\limits_{k = 1}^K {\sqrt {{\mu _k}} \xi _k^H{\bf{w}}_k^H{{\bf{H}}_{b,k}}{{\bf{F}}_{{\rm{RF,}}b}}{{\bf{f}}_{{\rm{BB,}}b,k}}} } \right) + D_{ b,k,j},
\end{align}
where ${C_{b,k,j}}$ and $D_{b,k,j}$  are irrelevant terms for $b$th BS, which can be respectively given by
\begin{equation}\label{SecIIIHB07}
\begin{aligned}
{C_{b,k,j}} = \sum\limits_{p \ne b}^B {{\bf{w}}_k^H{{\bf{H}}_{p,k}}{{\bf{F}}_{{\rm{RF,}}p}}{{\bf{f}}_{{\rm{BB,}}p,j}}},
\;\; D_{b,k,j} =  - 2\Re \left( {\sum\limits_{k = 1}^K {\sum\limits_{p \ne b}^B {\sqrt {{\mu _k}} \xi _k^H{\bf{w}}_k^H{{\bf{H}}_{p,k}}{{\bf{F}}_{{\rm{RF,}}p}}{{\bf{f}}_{{\rm{BB,}}p,j}}} } } \right).
\end{aligned}
\end{equation}

To solve the constrained problem (\ref{SecIIIHB04}), ADMM is an efficient tool that blends the benefits of dual decomposition and augmented Lagrangian method \cite{System02}. By considering ${{\bf{F}}_b} = {{\bf{F}}_{{\rm{RF,}}b}}{{\bf{F}}_{{\rm{BB,}}b}}$ into problem (\ref{SecIIIHB04}), the augmented Lagrangian function can be formulated as
\begin{align}\label{SecIIIHB08}
{{\cal L}_1}  \left( {{{\bf{F}}_b},{{\bf{F}}_{{\rm{RF,}}b}},{{\bf{F}}_{{\rm{BB,}}b}},{\bf{\Delta }}} \right) = {f_1}\left( {{{\bf{F}}_{{\rm{RF,}}b}},{{\bf{F}}_{{\rm{BB,}}b}}} \right)+ \frac{{{\rho _1}}}{2}\left\| {{{\bf{F}}_b} - {{\bf{F}}_{{\rm{RF,}}b}}{{\bf{F}}_{{\rm{BB,}}b}} + \frac{{\bf{\Delta }}}{{{\rho _1}}}} \right\|_F^2 + {\bf E}\left( {\bf{\Delta }} \right),
\end{align}
where ${\rho _1} > 0$ indicate the dual variable and the penalty parameter and ${\bf E}\left( {\bf{\Delta }} \right) =  - {\rm{tr(}}{{\bf{\Delta }}^H}{\bf{\Delta }}{\rm{)}}/2{\rho _1}$  is a constant term when ${\bf{\Delta }} \in {{\mathbb C}^{{N_t} \times K}}$  remains fixed.

Based on the working principle of ADMM \cite{System020,System021}, the iterative process for optimizing ${{\cal L}_1}\left( {{{\bf{F}}_b},{{\bf{F}}_{{\rm{RF,}}b}},{{\bf{F}}_{{\rm{BB,}}b}},{\bf{\Delta }}} \right)$ can be expressed as
\begin{subequations}\label{SecIIIHB09}
\begin{align}
&{\bf{F}}_b^{t+1} = \mathop {\arg \min}\limits_{{{\bf{F}}_b}}{{\cal L}_1}\left({{{\bf{F}}_b},{\bf{F}}_{{\rm{RF,}}b}^t,{\bf{F}}_{{\rm{BB,}}b}^t,{{\bf{\Delta }}^t}} \right) \notag \\ &\quad\quad\quad \; {\rm{s.t.}} \;\; \left\| {{{\bf{F}}_b}} \right\|_F^2 \le {P_b},\label{SecIIIHB09a}\\
&{\bf{F}}_{{\rm{RF,}}b}^{t + 1} = \mathop {\arg \min }\limits_{{{\bf{F}}_{{\rm{RF,}}b}}} {{\cal L}_1}\left( {{\bf{F}}_b^{t + 1}, {{\bf{F}}_{{\rm{RF,}}b}}, {\bf{F}}_{{\rm{BB,}}b}^t, {{\bf{\Delta }}^t}} \right) \notag \\
&\quad\quad\quad \;\; {\rm{s.t.}} \;\; {\left| {{{\bf{F}}_{{\rm{RF,}}b}}\left( {{i_1},{j_1}} \right)} \right|^2} = 1,\forall {i_1},{j_1},\label{SecIIIHB09b} \\
&{\bf{F}}_{{\rm{BB,}}b}^{t + 1} = \mathop {\arg \min }\limits_{{{\bf{F}}_{{\rm{BB,}}b}}} {{\cal L}_1}\left( {{\bf{F}}_b^{t + 1},{\bf{F}}_{{\rm{RF,}}b}^{t + 1},{{\bf{F}}_{{\rm{BB,}}b}},{{\bf{\Delta }}^t}} \right), \label{SecIIIHB09c} \\
&{{\bf{\Delta }}^{t + 1}} = {{\bf{\Delta }}^t} + {\rho _1}\left( {{\bf{F}}_b^{t + 1} - {\bf{F}}_{{\rm{RF,}}b}^{t + 1}{\bf{F}}_{{\rm{BB,}}b}^{t + 1}} \right), \label{SecIIIHB09d}
\end{align}
\end{subequations}
where superscript $t$ denotes the iteration index for proceeding ADMM algorithm.

\textbf{\emph{1)} Optimize} ${{\bf{F}}_b}$: Given fixed $({\bf{F}}_{{\rm{RF,}}b}^t,{\bf{F}}_{{\rm{BB,}}b}^t,{{\bf{\Delta }}^t})$, the minimization problem with regard to ${{\bf{F}}_b}$ can be written as
\begin{align}\label{SecIIIHB10}
& \mathop {\min }\limits_{{{\bf{F}}_b}} \frac{{{\rho _1}}}{2}\left\| {{{\bf{F}}_b} - {\bf{F}}_{{\rm{RF,}}b}^t{\bf{F}}_{{\rm{BB,}}b}^t + \frac{{{{\bf{\Delta }}^t}}}{{{\rho _1}}}} \right\|_F^2 \notag \\
& \; {\rm{s.t.}} \;\; \left\| {{{\bf{F}}_b}} \right\|_F^2 \le {P_b}.
\end{align}

Taking into account the power constraint, the augmented Lagrangian for problem (\ref{SecIIIHB10}) can be formulated as
\begin{align}\label{SecIIIHB11}
{{\cal L}_2}\left( {{{\bf{F}}_b},{\mu _1}} \right)  = \frac{{{\rho _1}}}{2}\left\| {{{\bf{F}}_b} - {\bf{F}}_{{\rm{RF,}}b}^t{\bf{F}}_{{\rm{BB,}}b}^t + \frac{{{{\bf{\Delta }}^t}}}{{{\rho _1}}}} \right\|_F^2 + {\tilde \mu _F}\left( {\left\| {{{\bf{F}}_b}} \right\|_F^2 - {P_b}} \right),
\end{align}
where ${\tilde \mu _F}$  indicates the Lagrange multiplier. By solving Karush-Kuhn-Tucker (KKT) conditions for (\ref{SecIIIHB11}), given by
\begin{equation}\label{SecIIIHB12}
\begin{aligned}
\frac{{\partial {{\cal L}_2}\left( {{{\bf{F}}_b},{{\tilde \mu }_F}} \right)}}{{\partial {{\bf{F}}_b}}} = 0, \;\;\; \left\| {{{\bf{F}}_b}} \right\|_F^2 \le {P_b},\;\;\;
{{\tilde \mu }_F}\left( {\left\| {{{\bf{F}}_b}} \right\|_F^2 - {P_b}} \right) = 0, \;\;\; {{\tilde \mu }_F} \ge 0,
\end{aligned}
\end{equation}
a closed-form solution to ${{\bf{F}}_b}$  can be expressed as
\begin{align}\label{SecIIIHB13}
{\bf{F}}_b^{t+1} = \frac{{\sqrt {{P_b}} \left( {{\rho _1}{\bf{F}}_{{\rm{RF,}}b}^t{\bf{F}}_{{\rm{BB,}}b}^t - {{\bf{\Delta }}^t}} \right)}}{{{{\left\| {{\rho _1}{\bf{F}}_{{\rm{RF,}}b}^t{\bf{F}}_{{\rm{BB,}}b}^t - {{\bf{\Delta }}^t}} \right\|}_F}}}.
\end{align}

\textbf{\emph{2)} Optimize} ${{\bf{F}}_{{\rm{RF,}}b}}$: With fixed  $({{\bf{F}}_b^{t + 1},{\bf{F}}_{{\rm{BB,}}b}^t,{{\bf{\Delta }}^t}})$, the optimization problem (\ref{SecIIIHB09b}) for ${{\bf{F}}_{{\rm{RF,}}b}}$  can be reformulated as
\begin{align}\label{SecIIIHB14}
& {\bf{F}}_{{\rm{RF,}}b}^{t + 1} =  \mathop {\arg \min }\limits_{{{\bf{F}}_{{\rm{RF,}}b}}} {f_2}\left( {{{\bf{F}}_{{\rm{RF,}}b}}} \right) \notag \\
&\quad \quad \quad \;\; {\rm{s.t.}} \;\; {\left| {{{\bf{F}}_{{\rm{RF,}}b}}\left( {{i_1},{j_1}} \right)} \right|^2} = 1,\forall {i_1},{j_1},
\end{align}
where
\begin{align}\label{SecIIIHB15}
{f_2}\left( {{{\bf{F}}_{{\rm{RF,}}b}}} \right)  & = \frac{{{\rho _1}}}{2}\left\| {{\bf{F}}_b^{t + 1} - {{\bf{F}}_{{\rm{RF,}}b}}{\bf{F}}_{{\rm{BB,}}b}^t + \frac{{{{\bf{\Delta }}^t}}}{{{\rho _1}}}} \right\|_F^2 - 2\Re \left( {\sum\limits_{k = 1}^K {\sqrt {{\mu _k}} \xi _k^H{\bf{w}}_k^H{{\bf{H}}_{b,k}}{{\bf{F}}_{{\rm{RF,}}b}}{\bf{f}}_{{\rm{BB,}}b,k}^t} } \right) \notag \\
& \quad\quad\quad\quad\quad\quad\quad\quad\quad\quad\quad\quad\quad + \sum\limits_{k = 1}^K {\sum\limits_{j = 1}^K {{{\left| {{\xi _k}} \right|}^2}{{\left| {{\bf{w}}_k^H{{\bf{H}}_{b,k}}{{\bf{F}}_{{\rm{RF,}}b}}{\bf{f}}_{{\rm{BB,}}b,j}^t + C_{ b,k,j}^t} \right|}^2}} }.
\end{align}

In order to leverage the MO method, we need to convert a constrained problem (\ref{SecIIIHB14}) into an unconstrained optimization problem over smooth Riemannian manifolds \cite{System02a}. Assuming that ${{\bf{f}}_{{\rm{RF}},b}} = {\rm{vec}}({{\bf{F}}_{{\rm{RF}},b}})$, problem (\ref{SecIIIHB14}) can be rewritten as
\begin{align}\label{SecIIIHB16}
{\bf{f}}_{{\rm{RF}},b}^{t + 1} = \mathop {\arg \min }\limits_{{{\bf{f}}_{{\rm{RF}},b}} \in {{\cal M}_b}} {\rm{ }}{f_3}\left( {{{\bf{f}}_{{\rm{RF}},b}}} \right)
\end{align}
where
\begin{align}\label{SecIIIHB17}
& {f_3}\left( {{{\bf{f}}_{{\rm{RF}},b}}} \right) = {\bf{f}}_{{\rm{RF}},b}^H\left( {\frac{{{\rho _1}}}{2}{\bf{B}}_b^{t,H}{\bf{B}}_b^t + {\bf{\Sigma }}_b^t} \right){{\bf{f}}_{{\rm{RF}},b}} + 2\Re \left({{\bf{\Pi }}_b^{t,H}}{{\bf{f}}_{{\rm{RF}},b}}\right), \; {\bf{a}}_{b,k,j}^t = {\bf{f}}_{{\rm{BB,}}b,j}^{t,T} \otimes \left( {{\bf{w}}_k^H{{\bf{H}}_{b,k}}} \right),  \notag\\
& {\bf{B}}_b^t = {\bf{F}}_{{\rm{BB,}}b}^{t,T} \otimes {{\bf{I}}_{{N_t}}}, \; {\bf{\Sigma }}_b^t = \sum\limits_{k = 1}^K {\sum\limits_{j = 1}^K {{{\left| {{\xi _k}} \right|}^2}{\bf{a}}_{b,k,j}^{t,H}{\bf{a}}_{b,k,j}^t} }, \; {\bf{b}}_b^t = \sum\limits_{k = 1}^K {\sqrt {{\mu _k}}{{\left( {\xi _k^H{\bf{f}}_{{\rm{BB,}}b,k}^{t,T} \otimes \left({{\bf{w}}_k^H{{\bf{H}}_{b,k}}} \right)} \right)}^H}}, \notag \\
& {\bf{\Pi }}_b^t = {\sum\limits_{k = 1}^K {\sum\limits_{j = 1}^K {{{\left| {{\xi _k}} \right|}^2}{\bf{a}}_{b,k,j}^{t,H}C_{b,k,j}^t} }  - \frac{{{\rho _1}}}{2}{\bf{B}}_b^{t,H} {\bf{m}}_b^t - {\bf{b}}_b^t}, \;\; {\bf{m}}_b^t = {\rm{vec}}\left( {{\bf{F}}_b^{t + 1}} \right) + \frac{{{\rm{vec}}\left( {{{\bf{\Delta }}^t}} \right)}}{{{\rho _1}}},
\end{align}
and the proof for problem (\ref{SecIIIHB16}) is provided in Appendix A.

In addition,  ${{\cal M}_b}$ represents the complex circle Riemannian manifold that can be specifically expressed as
\begin{align}\label{SecIIIHB19}
{{\cal M}_b} = \left\{ {{{\bf{f}}_{{\rm{RF}},b}} \in {{\mathbb C}^{{N_t}{N_{{\rm{RF}}}}}}:\left| {{{\bf{f}}_{{\rm{RF}},b}}} \right| = {{\bf{1}}_{{N_t}{N_{{\rm{RF}}}}}}} \right\}.
\end{align}

We note that the search space with ${N_t}{N_{{\rm{RF}}}}$  complex circles denotes the Riemannian submanifold of ${{\mathbb C}^ {{N_t} {N_{{\rm{RF}}}}}}$, and any point on this submanifold can be linearized by a tangent space. Given a fixed point ${{\bf{f}}_{{\rm{RF}},b}} \in {{\cal M}_b}$, the tangent space can be defined as
\begin{align}\label{SecIIIHB20}
{T_{{{\cal M}_b}}} = \left\{ {{\bf{z}} \in {{\mathbb C}^{{N_t}{N_{{\rm{RF}}}}}}:\Re \left\{ {{\bf{z}} \circ {{\bf{f}}_{{\rm{RF}},b}}} \right\} = {{\bf{0}}_{{N_t}{N_{{\rm{RF}}}}}}} \right\},
\end{align}
where ${\bf{z}}$  indicates the tangent vector at the point ${{\bf{f}}_{{\rm{RF}},b}} \in {{\cal M}_b}$. More directly, ${T_{{{\cal M}_b}}}$ can be further regarded as the set of all tangent vectors orthogonal to the given point, and one of these tangent vectors is the negative Riemannian gradient that implies the fastest descent direction \cite{System02b}. To this end, we define a unique tangent vector ${\rm{grad}}{f_3}({{\bf{f}}_{{\rm{RF}},b}})$ as the Riemannian gradient at ${{\bf{f}}_{{\rm{RF}},b}} \in {{\cal M}_b}$, given by the orthogonal projection of the Euclidean gradient to the tangent spaces as follow
\begin{align}\label{SecIIIHB21}
{\rm{grad}}{f_3}\left( {{{\bf{f}}_{{\rm{RF}},b}}} \right) = \nabla {f_3}\left( {{{\bf{f}}_{{\rm{RF}},b}}} \right) - \Re \left\{ {\nabla {f_3}\left( {{{\bf{f}}_{{\rm{RF}},b}}} \right) \circ {\bf{f}}_{{\rm{RF}},b}^*} \right\} \circ {{\bf{f}}_{{\rm{RF}},b}},
\end{align}
where
\begin{equation}\label{SecIIIHB22}
\begin{aligned}
\nabla {f_3}\left( {{{\bf{f}}_{{\rm{RF}},b}}} \right) = \left( {\frac{{{\rho _1}}}{2}{\bf{B}}_b^{t,H}{\bf{B}}_b^t + {\bf{\Sigma }}_b^t} \right){{\bf{f}}_{{\rm{RF}},b}} + {\bf{\Pi }}_b^t. \notag
\end{aligned}
\end{equation}

To guarantee that ${\rm{grad}}{f_3}({{\bf{f}}_{{\rm{RF}},b}})$ is mapped from ${T_{{{\cal M}_b}}}$ onto the manifold itself, the retraction operation is proposed to determine the following point remaining on the prescribed manifold while moving along a tangent vector. Thus, the retraction of  ${\rm{grad}}{f_3}({{\bf{f}}_{{\rm{RF}},b}})$ at ${{\bf{f}}_{{\rm{RF}},b}} \in {{\cal M}_b}$ can be expressed as
\begin{align}\label{SecIIIHB21}
{\bf{f}}_{{\rm{RF}},b}^{i + 1}  = {{\mathop{\rm Retr}\nolimits} _{{\bf{f}}_{{\rm{RF}},b}^i}}\left( { - {\ell ^i}{\rm{grad}}{f_3}\left( {{\bf{f}}_{{\rm{RF}},b}^i} \right)} \right)
= {\rm{vec}}\left( {\frac{{{\bf{f}}_{{\rm{RF}},b}^i - {\ell ^i}{\rm{grad}}{f_3}\left( {{\bf{f}}_{{\rm{RF}},b}^i} \right)}}{{\left| {{\bf{f}}_{{\rm{RF}},b}^i - {\ell^i}{\rm{grad}}{f_3}\left( {{\bf{f}}_{{\rm{RF}},b}^i} \right)} \right|}}} \right),
\end{align}
where superscript $i$ denotes the iteration index, ${\ell^i}$ is the step-size, and ${{\mathop{\rm Retr}\nolimits}}\left(  \cdot  \right)$ is the retraction operation from ${T_{{{\cal M}_b}}}$ to ${{\cal M}_b}$, respectively. Particularly, the step-size selection strategy for the MO algorithm is Armijo backtracking line search. Based on the above analysis, problem (\ref{SecIIIHB16}) can be well solved by the MO method .

\textbf{\emph{3)} Optimize} ${{\bf{F}}_{{\rm{BB,}}b}}$: Given the set $({\bf{F}}_b^{t + 1},{\bf{F}}_{{\rm{RF,}}b}^{t + 1},{{\bf{\Delta }}^t})$, the DBF design ${{\bf{F}}_{{\rm{BB,}}b}}$ for problem (\ref{SecIIIHB09c}) can be expressed as
\begin{align}\label{SecIIIHB24}
{\bf{F}}_{{\rm{BB,}}b}^{t + 1} = \mathop {\arg \min }\limits_{{{\bf{F}}_{{\rm{BB,}}b}}} {f_2}\left( {{{\bf{F}}_{{\rm{BB,}}b}}} \right),
\end{align}
where ${f_2}\left( {{{\bf{F}}_{{\rm{BB,}}b}}} \right)$ is presented in (\ref{SecIIIHB15}). By defining ${{\bf{f}}_{{\rm{BB,}}b}} = {\rm{vec}}({{\bf{F}}_{{\rm{BB,}}b}})$, problem (\ref{SecIIIHB24}) can be reformulated as
\begin{align}\label{SecIIIHB26}
{\bf{f}}_{{\rm{BB,}}b}^{t + 1} = \mathop {\arg \min }\limits_{{{\bf{f}}_{{\rm{BB,}}b}}} {f_4}\left( {{{\bf{f}}_{{\rm{BB,}}b}}} \right)
\end{align}
where
\begin{align}\label{SecIIIHB27}
& {f_4}\left( {{{\bf{f}}_{{\rm{BB,}}b}}} \right) = {\bf{f}}_{{\rm{BB,}}b}^H\left( {{\bf{C}}_b^t + \frac{{{\rho _1}}}{2}{\bf{D}}_b^{t,H}{\bf{D}}_b^t} \right){{\bf{f}}_{{\rm{BB,}}b}}
+ 2\Re\left( \left( {{\bf{r}}_b^{t,T} + {\bf{d}}_b^{t,H} - \frac{{{\rho _1}}}{2}{\bf{m}}_b^{t,H}{\bf{D}}_b^t} \right){{\bf{f}}_{{\rm{BB,}}b}}\right),\notag\\
& {\bf{\tilde d}}_{b,j}^t = \sum\limits_{k = 1}^K {{{\left| {{\xi _k}} \right|}^2}{\bf{c}}_{b,k}^{t, * }C_{\bar b,k,j}^t}, \; {\bf{d}}_b^t = {[ {{\bf{\tilde d}}_{b,1}^{t,T}, \cdots ,{\bf{\tilde d}}_{b,K}^{t,T}} ]^T}, \; {\bf{c}}_{b,k}^{t,T} = {\bf{w}}_k^H{{\bf{H}}_{b,k}}{\bf{F}}_{{\rm{RF,}}b}^{t + 1}, \; {\bf{D}}_b^t = {{\bf{I}}_K} \otimes {\bf{F}}_{{\rm{RF,}}b}^{t + 1}, \notag\\
& {\bf{C}}_b^t = {{\bf{I}}_K} \otimes ( {\sum\limits_{k = 1}^K {{{\left| {{\xi _k}} \right|}^2}{\bf{c}}_{b,k}^{t, * }{\bf{c}}_{b,k}^{t,T}}}), \; {\bf{\tilde r}}_{b,k}^{t,T} =  - \sqrt {{\mu _k}} \xi _k^H{\bf{w}}_k^H{{\bf{H}}_{b,k}}{\bf{F}}_{{\rm{RF,}}b}^{t + 1}, \; {\bf{r}}_b^t = {[ {{\bf{\tilde r}}_{b,1}^{t,T},\cdots ,{\bf{\tilde r}}_{b,K}^{t,T}} ]^T}.
\end{align}

Since problem (\ref{SecIIIHB26}) with quadratic objective function is convex and unconstrained, the optimal solution can be obtained directly by setting $\partial {f_4}({{\bf{f}}_{{\rm{BB,}}b}})/\partial {{\bf{f}}_{{\rm{BB,}}b}} = {\bf{0}}$. Thus, ${\bf{f}}_{{\rm{BB,}}b}^{t + 1}$ can be given by
\begin{align}\label{SecIIIHB29}
{\bf{f}}_{{\rm{BB,}}b}^{t + 1} = {\left( {{\bf{C}}_b^t + \frac{{{\rho _1}}}{2}{\bf{D}}_b^{t,H}{\bf{D}}_b^t} \right)^{ - 1}} \left( {\frac{{{\rho _1}}}{2}{\bf{D}}_b^{t,H}{\bf{m}}_b^t - {\bf{r}}_b^{t, * } - {\bf{d}}_b^t} \right).
\end{align}

\subsection{Fix \protect\boldmath$\left( {{\bf{F}},{\bf{\Phi }},{\bf{\lambda }},{\bf{\xi }}} \right)$  and Solve ${\bf{w}}$}
To design the combining vectors for $K$ users, the original problem (\ref{CFsystem14}) with regard to ${\bf{w}}$ can be rewritten as
\begin{align}\label{SecIIIHB30}
&\mathop {{\rm{min}}}\limits_{\bf{w}} \; {f_5}\left( {\bf{w}} \right) \notag \\
&\; {\rm{s.t.}} \;\; {\left| {{\bf{w}}\left( i \right)} \right|^2} = 1,\forall i = 1,2, \cdots ,K{N_r},
\end{align}
where
\begin{align}\label{SecIIIHB31}
{f_5}\left( {\bf{w}} \right) = \sum\limits_{k = 1}^K {\sum\limits_{j = 1}^K {{{| {{\xi _k}} |}^2}{{\left| {\sum\limits_{b = 1}^B {{\bf{w}}_k^H{{\bf{H}}_{b,k}}{{\bf{F}}_{{\rm{RF,}}b}}{{\bf{f}}_{{\rm{BB,}}b,j}}} } \right|}^2}} } - \Re \left( {\sum\limits_{k = 1}^K {\sum\limits_{b = 1}^B {2\sqrt {{\mu _k}} \xi _k^H{\bf{w}}_k^H{{\bf{H}}_{b,k}}{{\bf{F}}_{{\rm{RF,}}b}}{{\bf{f}}_{{\rm{BB,}}b,k}}}}}\right).
\end{align}

After some algebraic transformations, ${f_5}\left( {\bf{w}} \right)$ can be simplified as a more compact form. Consequently, problem (\ref{SecIIIHB30}) can be equivalently written as
\begin{align}\label{SecIIIHB32}
& \mathop {{\rm{min}}}\limits_{\bf{w}} \; {{\bf{w}}^H}{\bf{PW}} - \Re \left( {2{{\bf{w}}^H}{\bf{q}}} \right) \notag \\
&\; {\rm{s.t.}} \;\; {\left| {{\bf{w}}\left( i \right)} \right|^2} = 1,\forall i = 1,2, \cdots ,K{N_r},
\end{align}
where
\begin{align}\label{SecIIIHB33}
& {{\bf{\tilde q}}_k} = \sum\limits_{b = 1}^B {\sqrt {{\mu _k}} \xi _k^H{{\bf{H}}_{b,k}}{{\bf{F}}_{{\rm{RF,}}b}}{{\bf{f}}_{{\rm{BB,}}b,k}}}, \; {{\bf{p}}_{k,j}} = \sum\limits_{b = 1}^B {{{\bf{H}}_{b,k}}{{\bf{F}}_{{\rm{RF,}}b}}{{\bf{f}}_{{\rm{BB,}}b,j}}}, \; {{\bf{\tilde P}}_k} = \sum\limits_{j = 1}^K {{{\left| {{\xi _k}} \right|}^2}{{\bf{p}}_{k,j}}{\bf{p}}_{k,j}^H}, \notag \\
& {\bf{q}} = {\left[ {{\bf{\tilde q}}_1^T,{\bf{\tilde q}}_2^T, \cdots ,{\bf{\tilde q}}_K^T} \right]^T}, \; {\bf{P}} = {\rm{diag}}( {{{{\bf{\tilde P}}}_1},{{{\bf{\tilde P}}}_2}, \cdots ,{{{\bf{\tilde P}}}_K}}).
\end{align}

Notably, problem (\ref{SecIIIHB32}) is a non-convex QCQP problem with quadratic objective function and unit-modulus constraints, which is similar to problem (\ref{SecIIIHB16}). In this regard, the MO algorithm can be performed to seek a high-quality solution.

\subsection{Fix \protect\boldmath$\left( {{\bf{F}},{\bf{w}},{\bf{\lambda }},{\bf{\xi }}} \right)$  and Solve ${\bf{\Phi}}$}
Given the optimized set  $({\bf{F}},{\bf{w}},{\bm{\lambda }},{\bm{\xi }})$, the remaining compound terms in (\ref{CFsystem13}) is only related to $\bf{\Phi}$. Hence,the passive BF design for $R$ RISs can be written as
\begin{align}\label{SecIIIHB38}
& \mathop {{\rm{min}}}\limits_{\bf{\Phi }} {\rm{  }}{f_6}\left( {\bf{\Phi }} \right) \notag \\
&\;{\rm{s.t.}} \;\;{\left| {{\bf{\Phi }}\left( {{i_2},{i_2}} \right)} \right|^2} \le 1,\forall {i_2} = 1,2, \cdots ,RM,
\end{align}
where
\begin{equation}\label{SecIIIHB39}
\begin{aligned}
{f_6}\left( {\bf{\Phi }} \right) =  & - \Re \left( {\sum\limits_{k = 1}^K {\sum\limits_{b = 1}^B {2\sqrt {{\mu _k}} \xi _k^H{\bf{w}}_k^H{{\bf{V}}_k}{\bf{\Phi }}{{\bf{G}}_b}{{\bf{F}}_{{\rm{RF,}}b}}{{\bf{f}}_{{\rm{BB,}}b,k}}} } } \right) \\
& + \sum\limits_{k = 1}^K {\sum\limits_{j = 1}^K {{{\left| {{\xi _k}} \right|}^2}{{\left| {\sum\limits_{b = 1}^B {{\bf{w}}_k^H{{\bf{V}}_k}{\bf{\Phi }}{{\bf{G}}_b}{{\bf{F}}_{{\rm{RF,}}b}}{{\bf{f}}_{{\rm{BB,}}b,j}}}  + E_{k,j}} \right|}^2}} },
\end{aligned}
\end{equation}
and ${E_{k,j}} = \sum\limits_{b = 1}^B {{\bf{w}}_k^H{{{\bf{\bar H}}}_{b,k}}{{\bf{F}}_{{\rm{RF,}}b}}{{\bf{f}}_{{\rm{BB,}}b,j}}}$.

Base on problem (\ref{SecIIIHB38}), ${f_6}( {\bf{\Phi }} )$ is still laborious to be handled. By introducing ${\bm{\varphi }} = {\bf{\Phi }}{{\bf{1}}_{RM}}$, problem (\ref{SecIIIHB38}) can be further simplified as
\begin{align}\label{SecIIIHB40}
& \mathop {{\rm{min}}}\limits_{\bm{\varphi }} \; {{\bm{\varphi }}^H}{\bf{Z}}{\bm{\varphi }}  + {{\bm{\varphi }}^H}{\bm{\kappa }} + {{\bm{\kappa}}^H}{\bm{\varphi}} \notag \\
&\; {\rm{s.t.}} \;\; {\left| {{\bm{\varphi }}\left( {{i_2}} \right)} \right|^2} \le 1, \; \forall {i_2} = 1,2, \cdots ,RM,
\end{align}
where
\begin{align}\label{SecIIIHB41}
& {{\bf{u}}_{b,k}} = \sqrt {{\mu _k}} \xi _k^H {\rm{diag}}\left( {{\bf{w}}_k^H{{\bf{V}}_k}} \right){{\bf{G}}_b}{{\bf{F}}_{{\rm{RF,}}b}}{{\bf{f}}_{{\rm{BB,}}b,k}}, \;\; {\bm{\kappa }} = \sum\limits_{k = 1}^K {\sum\limits_{j = 1}^K {\sum\limits_{b = 1}^B {{{\left| {{\xi _k}} \right|}^2}{E_{k,j}}{\bf{v}}_{b,k,j}^ * } - \sum\limits_{k = 1}^K {\sum\limits_{b = 1}^B {{\bf{u}}_{b,k}^ * } } } }, \notag \\
& {{\bf{v}}_{b,k,j}} = {\rm{diag}}\left( {{\bf{w}}_k^H{{\bf{V}}_k}} \right){{\bf{G}}_b}{{\bf{F}}_{{\rm{RF,}}b}}{{\bf{f}}_{{\rm{BB,}}b,j}}, \;\; {\bf{Z}} = {\sum\limits_{k = 1}^K {\sum\limits_{j = 1}^K {{{\left| {{\xi _k}} \right|}^2}\left( {\sum\limits_{b = 1}^B {{\bf{v}}_{b,k,j}^ * } } \right)\left( {\sum\limits_{b = 1}^B {{\bf{v}}_{b,k,j}^T} } \right)} } }.
\end{align}

\begin{algorithm}[t]
\caption{Proposed cooperative beamforming scheme.}\label{alg1}
\begin{algorithmic}
\STATE
\STATE \textbf{Input:} ${{\bf{\bar H}}_{b,k}}$, ${{\bf{G}}_{b,r}}$, ${{\bf{V}}_{r,k}}$, $b \in {\cal B}$, $r \in {\cal R}$, $k \in {\cal K}$.
\STATE \textbf{Initialization:} ${\bf{F}}_{{\rm{RF,}}b}$, ${\bf{F}}_{{\rm{BB,}}b}$, ${\bf{\Phi }}$, ${\bf{w}}$, ${\bf{\Delta }}$, ${\rho _1}$, $i=1$.

\STATE 1: \textbf{while} $i \le {I_{max}}$
\STATE 2: \hspace{0.2cm} Update $\lambda_k$ and $\xi_k$ according to (\ref{SecIIIHB01});
\STATE 4: \hspace{0.2cm} \textbf{for} $b=1:B$ \textbf{in parallel}
\STATE 5: \hspace{0.2cm} \hspace{0.2cm} \textbf{for} $t=1:t_{max}$

\STATE 6: \hspace{0.2cm} \hspace{0.2cm} \hspace{0.15cm} Update $\left({\bf{F}}_b^{t + 1}, {\bf{F}}_{{\rm{RF,}}b}^{t + 1}, {\bf{F}}_{{\rm{BB,}}b}^{t + 1}, {\bf{\Delta }}^{t + 1} \right)$ by solving (\ref{SecIIIHB09a})-(\ref{SecIIIHB09d});

\STATE 7: \hspace{0.2cm} \hspace{0.2cm} \textbf{end for}
\STATE 8: \hspace{0.2cm} \textbf{end for}
\STATE 9: \hspace{0.2cm} Update ${\bf{w}}$ and ${\bf{\Phi}}$ by solving (\ref{SecIIIHB32}) and (\ref{SecIIIHB40});
\STATE 10: \textbf{end while}
\STATE 11: Calculate $R_{sum}$ according to (\ref{CFsystem10});

\STATE \textbf{Output:} ${\bf{F}}_{{\rm{RF,}}b}^{\star}$, ${\bf{F}}_{{\rm{BB,}}b}^{\star}$, ${\bf{w}}^{\star}$, ${\bf{\Phi}}^{\star}$, $R_{sum}$.

\end{algorithmic}
\label{alg1}
\end{algorithm}

To solve the convex problem (\ref{SecIIIHB40}) with convex constraints, the PDS method can be leveraged to get a desired solution in a computation-efficient way \cite{introduction13, System03}. Furthermore, we continue the iterative process by optimizing the variable set $\left( {{\bf{F}},{\bf{\Phi}},{\bf{w}},{\bm{\lambda}},{\bm{\xi}}} \right)$ until the AO algorithm converges. The overall procedure for CBF design is summarized in Algorithm \ref{alg1}.

\section{Base Station Selection for RIS-Aided P-CF-mMIMO System}
The fully-connected CF-mMIMO architecture results in inevitably extravagant network costs due to a vast number of communication links. To suppress resultant communication costs, we develop a novel P-CF-mMIMO network architecture that selects partial communication links among BSs and users in this section. Concretely, the CPU is able to determine whether one BS can connect each user based on channel conditions. Moreover, we introduce the integer programming to solve the BS selection problem with binary integer constraints so as to assist the CPU in making decisions.

\subsection{Problem Formulation for BS Selection}

With respect to the RIS-aided P-CF-mMIMO system, the transmitted signal from $b$th BS can be expressed as
\begin{align}\label{SecIVBS01}
{{\bf{\hat x}}_b} = {{\bf{F}}_{{\rm{RF,}}b}}{{\bf{F}}_{{\rm{BB,}}b}}{{\bf{{\Lambda}}}_b}{\bf{s}} = \sum\limits_{k = 1}^K {{{\bf{F}}_{{\rm{RF,}}b}}{{\bf{f}}_{{\rm{BB,}}b,k}}{\tau _{b,k}}{s_k}},
\end{align}
where ${{\bf{{\Lambda}}}_b} = {\rm{diag}}({\tau _{b,1}},{\tau _{b,2}}, \cdots ,{\tau _{b,K}})$ is a diagonal matrix determining network connections among users and $b$th BS and the $k$th entry satisfies ${\tau _{b,k}} \in \{ 0,1\}$. More precisely, ${\tau _{b,k}} =1$ if $b$th BS communicates with $k$th user, otherwise ${\tau _{b,k}} =0$, which can be dynamically controlled by the CPU. We define ${{\cal K}_b} = \{ 1,2, \cdots ,{K_b}\}$ $(\forall b \in {\cal B})$ as the subset of users that is served by $b$th BS. Similarly, ${{\cal B}_k} = \{ 1,2, \cdots ,{B_k}\}$ $(\forall k \in {\cal K})$ is defined as the subset of BSs connecting to $k$th user. Based on the number of connected links between users and BSs, we can further get the condition of $\sum\nolimits_{b \in {\cal B}} {{K_b}}  = \sum\nolimits_{k \in {\cal K}} {{B_k}}$.

To better evaluate the influence brought by the number of communication links, we define a new metric for the proposed P-CF-mMIMO system as network connection ratio (NCR), which can be further expressed as
\begin{align}\label{SecIVBS02}
{\alpha } = \frac{{\sum\nolimits_{b \in {\cal B}} {{K_b}} }}{{BK}},
\end{align}
where ${\alpha} \in \left( {0,1} \right)$. This metric plays a significant role in characterizing communication costs, including the integration of latency, computational complexity, signaling overhead, backhaul overhead and energy consumption. In fact, NCR determines the total number of communication links among BSs and users. For instance, when ${\alpha}$ is smaller, the communication costs decline as well as WSR performance. Given larger ${\alpha}$, both communication costs and WSR performance raise accordingly. Hence, by adjusting ${\alpha}$, the P-CF-mMIMO can make a better tradeoff between performance and communication costs.

For the RIS-aided P-CF-mMIMO network, the SINR of $k$th user can be calculated as
\begin{align}\label{SecIVBS03}
{\hat \gamma _k} = \frac{{{{\left| {\sum\limits_{b = 1}^B {{\bf{w}}_k^H{{\bf{H}}_{b,k}}{{\bf{F}}_{{\rm{RF,}}b}}{{\bf{f}}_{{\rm{BB,}}b,k}}{\tau _{b,k}}} } \right|}^2}}}{{\sum\limits_{j = 1,j \ne k}^K {{{\left| {\sum\limits_{b = 1}^B {{\bf{w}}_k^H{{\bf{H}}_{b,k}}{{\bf{F}}_{{\rm{RF,}}b}}{{\bf{f}}_{{\rm{BB,}}b,j}}{\tau _{b,j}}} } \right|}^2}}  + {\sigma ^2}}}.
\end{align}

Then, the WSR for $K$ users can be written as
\begin{align}\label{SecIVBS04}
{\hat R_{sum}} = \sum\limits_{k = 1}^K {{\omega _k}\log \left( {1 + {{\hat \gamma }_k}} \right)}.
\end{align}


Similar to the previous section, by using the FP technique \cite{System01},  the objective function (\ref{SecIVBS04}) of the WSR optimization problem can be rewritten as
\begin{align}\label{SecIVBS05}
\hat f\left( {{\bf{\hat F}},{\bf{\hat \Phi }},{\bf{\hat w}},{\bm{\hat \lambda }},{\bm{\hat \xi }}} \right) = - \sum\limits_{k = 1}^K {{\omega _k}\log \left( {1 + {\lambda _k}} \right)} & - \Re \left( {\sum\limits_{k = 1}^K {\sum\limits_{b = 1}^B {2\sqrt {{\mu _k}} \xi _k^H{\bf{w}}_k^H{{\bf{H}}_{b,k}}{{\bf{F}}_{{\rm{RF,}}b}}{{\bf{f}}_{{\rm{BB,}}b,k}}{\tau _{b,k}}} } } \right) \notag \\
+ \sum\limits_{k = 1}^K {{\omega _k}{\lambda _k}} + \sum\limits_{k = 1}^K {{{\left| {{\xi _k}} \right|}^2}{\sigma ^2}} & + \sum\limits_{k = 1}^K {\sum\limits_{j = 1}^K {{{\left| {{\xi _k}} \right|}^2} {{\left| {\sum\limits_{b = 1}^B {{\bf{w}}_k^H{{\bf{H}}_{b,k}}{{\bf{F}}_{{\rm{RF,}}b}}{{\bf{f}}_{{\rm{BB,}}b,j}}{\tau _{b,j}}} } \right|}^2}} },
\end{align}

Based on the transformed function (\ref{SecIVBS05}), the joint optimization of CBF design and BS selection for RIS-aided P-CF-mMIMO systems can be formulated as
\begin{subequations}\label{SecIVBS06}
\begin{align}
& \mathop {{\rm{min}}}\limits_{{\bf{\hat F}},{\bf{\hat \Phi }},{\bf{\hat w}},{\bf{{\Lambda}}}} \; \hat f\left( {{\bf{\hat F}},{\bf{\hat \Phi }},{\bf{\hat w}},{\bm{\hat \lambda }},{\bm{\hat \xi }}} \right), \label{SecIVBS06main} \\
& \;\;\;{\rm{s.t.}} \; \left\| {{{\bf{F}}_{{\rm{RF,}}b}}{{\bf{F}}_{{\rm{BB,}}b}}{{\bf{{\Lambda}}}_b}} \right\|_F^2 \le {P_b},\forall b \in {\cal B},\label{SecIVBS06a}\\
& \;\;\;\;\;\;\;\;\;\; {\left| {{\bf{F}}_{{\rm{RF}},b}\left( {{i_1},{j_1}} \right)} \right|^2} = 1,\forall {i_1},{j_1},b \in {\cal B},\label{SecIVBS06b}\\
& \;\;\;\;\;\;\;\;\;\; {\left| {{\bf{\hat \Phi }}\left( {{i_2},{i_2}} \right)} \right|^2} \le 1,\forall {i_2}, \label{SecIVBS06c} \\
& \;\;\;\;\;\;\;\;\;\; {\left| {{\bf{\hat w}}\left( {{i_3}} \right)} \right|^2} = 1,\forall {i_3}, \label{SecIVBS06d} \\
& \;\;\;\;\;\;\;\;\;\; {{\bf{{\Lambda}}}_b}({i_4},{i_4}) \in \left\{ {0,1} \right\},\forall {i_4},b \in {\cal B}, \label{SecIVBS06e}
\end{align}
\end{subequations}
where ${\bf{{\Lambda}}} = \{ {{\bf{{\Lambda}}}_b}|\;\forall b \in {\cal B}\}$, ${\bf{\hat F}} = \{ {{\bf{\hat F}}_b}|\;\forall b \in {\cal B}\}$ and ${{\bf{\hat F}}_b} = {{\bf{F}}_{{\rm{RF,}}b}}{{\bf{F}}_{{\rm{BB,}}b}}{{\bf{{\Lambda}}}_b}$,. The main difference between problem (\ref{CFsystem14}) and problem (\ref{SecIVBS06}) is extra integer constraints (\ref{SecIVBS06a}) and (\ref{SecIVBS06e}). Hence, we pay more attention to the BS selection problem that incorporates integer constraints.

\subsection{Fix $( {{\bf{\hat F}},{\bf{\hat \Phi }},{\bf{\hat w}}})$ and Solve \protect\boldmath $( {{\bf{\hat \lambda }},{\bf{\hat \xi}}})$}
Similar to (\ref{SecIIIHB01}), the auxiliary variables $( {{{\hat \lambda }_k},{{\hat \xi }}_k})$ for $k$th user can be respectively given by
\begin{subequations}\label{SecIVBS07}
\begin{align}
{\hat \lambda} _k^{\star} = {{\hat \gamma} _k}, \;\;\;\;\; \hat \xi _k^{\star} = \frac{{\sqrt {{\mu _k}} \left( {\sum\limits_{b = 1}^B {{\bf{w}}_k^H{{\bf{H}}_{b,k}}{{\bf{F}}_{{\rm{RF,}}b}}{{\bf{f}}_{{\rm{BB,}}b,k}}{\tau _{b,k}}} } \right)}}{{\sum\limits_{j = 1}^K {\left( {{{\left| {\sum\limits_{b = 1}^B {{\bf{w}}_k^H{{\bf{H}}_{b,k}}{{\bf{F}}_{{\rm{RF,}}b}}{{\bf{f}}_{{\rm{BB,}}b,j}}{\tau _{b,j}}} } \right|}^2}} \right)}  + {\sigma ^2}}}, \;\; \forall k \in {\cal K}.\label{SecIVBS07b}
\end{align}
\end{subequations}

\subsection{Fix \protect\boldmath$({\bf{\hat \Phi }},{\bf{\hat w}},{\bf{\hat \lambda }},{\bf{\hat \xi }})$  and Solve ${\bf{\hat F}}$}
According to the P-CF-MIMO system, the BS selection matrix ${{\bf{{\Lambda}}}_b}$ influences the power allocation and the HBF design. Thereby, the HBF design at $b$th BS can be formulated as
\begin{subequations}\label{SecIVBS09}
\begin{align}
& \mathop {{\rm{min}}}\limits_{{{\bf{F}}_{{\rm{RF,}}b}},{{\bf{F}}_{{\rm{BB,}}b}}} \; {{\hat f}_1}\left( {{{\bf{F}}_{{\rm{RF,}}b}},{{\bf{F}}_{{\rm{BB,}}b}},{{\bf{{\Lambda}}}_b}} \right) \label{SecIVBS09main}\\
& \;\;\;\;\; {\rm{s.t.}} \;\; {\text{(\ref{SecIVBS06a})}},\;{\text{(\ref{SecIVBS06b})}},\;{\text{(\ref{SecIVBS06e})}},
\end{align}
\end{subequations}
where
\begin{align}\label{SecIVBS10}
{{\hat f}_1}\left( {{{\bf{F}}_{{\rm{RF,}}b}},{{\bf{F}}_{{\rm{BB,}}b}},{{\bf{A}}_b}} \right) = & - 2\Re \left( {\sum\limits_{k = 1}^K {\sqrt {{\mu _k}} \xi _k^H{\bf{w}}_k^H{{\bf{H}}_{b,k}}{{\bf{F}}_{{\rm{RF,}}b}}{{\bf{f}}_{{\rm{BB,}}b,k}}{\tau _{b,k}}} } \right) \notag \\
& +\sum\limits_{k = 1}^K {\sum\limits_{j = 1}^K {{{\left| {{\xi _k}} \right|}^2}{{\left| {{\bf{w}}_k^H{{\bf{H}}_{b,k}}{{\bf{F}}_{{\rm{RF,}}b}}{{\bf{f}}_{{\rm{BB,}}b,j}}{\tau _{b,j}} + {{\hat C}_{b,k,j}}} \right|}^2}} } + {{\hat D}_{b,k,j}},
\end{align}
and ${{\hat C}_{b,k,j}}$ and ${{\hat D}_{b,k,j}}$ are irrelevant terms for optimizing ${{\bf{F}}_{{\rm{RF,}}b}}$ and ${{\bf{F}}_{{\rm{BB,}}b}}$.

By considering an auxiliary variable ${{\bf{\hat F}}_b} = {{\bf{F}}_{{\rm{RF,}}b}}{{\bf{F}}_{{\rm{BB,}}b}}{{\bf{{\Lambda}}}_b}$, the augmented Lagrangian function for problem (\ref{SecIVBS09}) can be formulated as
\begin{align}\label{SecIVBS11}
{{\cal L}_3}\left( {{{{\bf{\hat F}}}_b},{{\bf{F}}_{{\rm{RF,}}b}},{{\bf{F}}_{{\rm{BB,}}b}},{{\bf{{\Lambda}}}_b},{\bf{\Delta }}} \right) & = {{\hat f}_1}\left( {{{\bf{F}}_{{\rm{RF,}}b}},{{\bf{F}}_{{\rm{BB,}}b}},{{\bf{{\Lambda}}}_b}} \right) \notag \\
& + \frac{{{\rho _2}}}{2}\left\| {{{{\bf{\hat F}}}_b} - {{\bf{F}}_{{\rm{RF,}}b}}{{\bf{F}}_{{\rm{BB,}}b}}{{\bf{{\Lambda}}}_b} + \frac{{\bf{\Delta }}}{{{\rho _2}}}} \right\|_F^2 - \frac{{{\rm{tr(}}{{\bf{\Delta }}^H}{\bf{\Delta }}{\rm{)}}}}{{2{\rho _2}}}.
\end{align}

Based on (\ref{SecIVBS11}), the ADMM algorithm can be utilized to solve $\left( {{{{\bf{\hat F}}}_b},{{\bf{F}}_{{\rm{RF,}}b}},{{\bf{F}}_{{\rm{BB,}}b}},{{\bf{{\Lambda}}}_b}} \right)$ in an iterative manner, which can be specifically expressed as
\begin{subequations}\label{SecIVBS12}
\begin{align}
& {\bf{\hat F}}_b^{t + 1} = \mathop {\arg \min }\limits_{{{\bf{F}}_b}} {{\cal L}_3}\left( {{{{\bf{\hat F}}}_b},{\bf{F}}_{{\rm{RF,}}b}^t, {\bf{F}}_{{\rm{BB,}}b}^t, {\bf{{\Lambda}}}_b^t, {{\bf{\Delta }}^t}} \right) \notag \\
&\quad\quad\quad\; {\rm{s.t.}} \;\; || {{{{\bf{\hat F}}}_b}} ||_F^2 \le {P_b},\label{SecIVBS12a} \\
& {\bf{F}}_{{\rm{RF,}}b}^{t + 1} = \mathop {\arg \min }\limits_{{\bf{F}}_b^{{\rm{RF}}}} {{\cal L}_3}\left( {{\bf{\hat F}}_b^{t + 1},{{\bf{F}}_{{\rm{RF,}}b}}, {\bf{F}}_{{\rm{BB,}}b}^t,{\bf{{\Lambda}}}_b^t,{{\bf{\Delta }}^t}} \right) \notag \\
&\quad\quad\quad\;\;  {\rm{s.t.}} \;\; {\left| {{{\bf{F}}_{{\rm{RF,}}b}}\left( {{i_1},{j_1}} \right)} \right|^2} = 1,\forall {i_1},{j_1},\label{SecIVBS12b} \\
& {\bf{F}}_{{\rm{BB,}}b}^{t + 1} = \mathop {\arg \min }\limits_{{{\bf{F}}_{{\rm{BB,}}b}}} {{\cal L}_3}\left( {{\bf{\hat F}}_b^{t + 1},{\bf{F}}_{{\rm{RF,}}b}^{t + 1},{{\bf{F}}_{{\rm{BB,}}b}},{\bf{{\Lambda}}}_b^t,{{\bf{\Delta }}^t}} \right), \label{SecIVBS12c} \\
& {\bf{{\Lambda}}}_b^{t + 1} = \mathop {\arg \min }\limits_{{{\bf{{\Lambda}}}_b}} {{\cal L}_3}\left( {{\bf{\hat F}}_b^{t + 1},{\bf{F}}_{{\rm{RF,}}b}^{t + 1},{\bf{F}}_{{\rm{BB,}}b}^{t + 1},{{\bf{{\Lambda}}}_b},{{\bf{\Delta }}^t}} \right)\notag \\
&\quad\quad\quad\;{\rm{s.t.}} \;\; {{\bf{{\Lambda}}}_b}({i_4},{i_4}) \in \left\{ {0,1} \right\},\forall {i_4},\label{SecIVBS12d} \\
& {{\bf{\Delta }}^{t + 1}} = {{\bf{\Delta }}^t} + {\rho _2}\left( {{\bf{F}}_b^{t + 1} - {\bf{F}}_{{\rm{RF,}}b}^{t + 1}{\bf{F}}_{{\rm{BB,}}b}^{t + 1}{\bf{{\Lambda}}}_b^{t + 1}} \right).\label{SecIVBS12e}
\end{align}
\end{subequations}

The subproblems (\ref{SecIVBS12a})-(\ref{SecIVBS12c}) in the P-CF-MIMO system are similar to aforementioned subproblems (\ref{SecIIIHB09a})-(\ref{SecIIIHB09c}) existing in the CF-MIMO system. Hence, the detailed steps for optimizing $({\bf{\hat F}}_b^{t+1},{\bf{F}}_{{\rm{RF,}}b}^{t+1},{\bf{F}}_{{\rm{BB,}}b}^{t+1})$ are omitted for brevity, and we concentrate on the BS selection problem.

\subsection{Relaxed Linear Approximation Method}
Given the optimized $({\bf{\hat F}}_b^{t+1},{\bf{F}}_{{\rm{RF,}}b}^{t+1},{\bf{F}}_{{\rm{BB,}}b}^{t+1}, {\bm{\hat \lambda }},{\bm{\hat \xi }})$, the BS selection problem (\ref{SecIVBS12d}) related to ${\bf{{\Lambda}}}_b$ can be reformulated as
\begin{align}\label{SecIVBS13}
{\bf{{\Lambda}}}_b^{t + 1} = & \mathop {\arg \min }\limits_{{{\bf{{\Lambda}}}_b}} {{\hat f}_2}\left( {{{\bf{{\Lambda}}}_b}} \right) \notag \\
&\; {\rm{s.t.}} \;\; {{\bf{{\Lambda}}}_b}({i_4},{i_4}) \in \left\{ {0,1} \right\},\forall {i_4},
\end{align}
where
\begin{align}\label{SecIVBS14}
{{\hat f}_2}\left( {{{\bf{{\Lambda}}}_b}} \right) & = \frac{{{\rho _2}}}{2}\left\| {{\bf{\hat F}}_b^{t + 1} - {\bf{F}}_{{\rm{RF,}}b}^{t + 1}{\bf{F}}_{{\rm{BB,}}b}^{t + 1}{{\bf{{\Lambda}}}_b} + \frac{{{{\bf{\Delta }}^t}}}{{{\rho _2}}}} \right\|_F^2 - 2\Re \left( {\sum\limits_{k = 1}^K {\sqrt {{\mu _k}} \xi _k^H{\bf{w}}_k^H{{\bf{H}}_{b,k}}{\bf{F}}_{{\rm{RF,}}b}^{t + 1}{\bf{f}}_{{\rm{BB,}}b,k}^{t + 1}{\tau _{b,k}}} } \right)\notag \\
& \quad\quad\quad\quad\quad\quad\quad\quad\quad\quad\quad\quad + \sum\limits_{k = 1}^K {\sum\limits_{j = 1}^K {{{\left| {{\xi _k}} \right|}^2}{{\left| {{\bf{w}}_k^H{{\bf{H}}_{b,k}}{\bf{F}}_{{\rm{RF,}}b}^{t + 1}{\bf{f}}_{{\rm{BB,}}b,j}^{t + 1}{\tau _{b,j}} + \hat C_{b,k,j}^t} \right|}^2}} }. \quad
\end{align}

%
%

To simplify problem (\ref{SecIVBS13}), we define a new vector ${{\bm{\tau }}_b} = {{\bf{{\Lambda}}}_b}{{\bf{1}}_K}$. As a consequence, the original problem (\ref{SecIVBS13}) can be recast as
\begin{subequations}\label{SecIVBS17}
\begin{align}
\mathop {\min }\limits_{{{\bm{\tau }}_b}} \; & {\bm{\tau }}_b^T{{\bf{U}}_b^t}{{\bm{\tau }}_b} + {\bf{r}}_b^{t,T}{{\bm{\tau }}_b} \label{SecIVBS17main} \\
{\rm{s.t.}}\;\; & {{\bf{I}}_K}{{\bm{\tau }}_b} \ge {{\bf{z}}_b}, \label{SecIVBS17a}\\
& {\bf{1}}_K^T{{\bm{\tau }}_b} = {K_b}, \label{SecIVBS17b} \\
& {{\bm{\tau }}_b} \in {\left\{ {0,1} \right\}^K}, \label{SecIVBS17c}
\end{align}
\end{subequations}
where
\begin{align}\label{SecIVBS18}
& {{\bf{U}}_b^t} = {\bf{L}}_b^t + \frac{{{\rho _2}}}{2}{\bf{\hat M}}_b^{t,H}{\bf{\hat M}}_b^t, \; {\bf{r}}_b^{t,T} = 2\Re( {{\bf{l}}_b^{t,H} - {\bf{g}}_b^{t,T} - \frac{{{\rho _2}}}{2}{\bf{\hat m}}_b^{t,T}{\bf{\hat M}}_b^t}), \;{\bf{\hat M}}_b^t = {[ {\Re ( {{\bf{\tilde M}}_b^{t,T}}),\Im ( {{\bf{\tilde M}}_b^{t,T}})} ]^T}, \notag \\
& {{\bf{z}}_b} = {[{z_{b,1}}, \cdots ,{z_{b,K}},]^T}, \; {z_{b,k}} = 1 - \small{\sum\limits_{{p \in {{\cal B} \backslash b}}} {{\tau _{p,k}}}}, \; {\bf{\tilde M}}_b^t = \left[ { \cdots ,{{\bf{M}}_b^t}( {:,(k - 1)K + k}), \cdots } \right], k \in {\cal K}, \notag \\
& {\bf{\hat m}}_b^t = {[ {\Re( {{\bf{m}}_b^{t,T}}),\Im ( {{\bf{m}}_b^{t,T}} )} ]^T}, \;\; {\bf{M}}_b^t = {{\bf{I}}_K} \otimes ( {{\bf{F}}_{{\rm{RF,}}b}^{t + 1}{\bf{F}}_{{\rm{BB,}}b}^{t + 1}}), \;\; g_{b,k}^t = \sqrt {{\mu _k}} \xi _k^H{\bf{w}}_k^H{{\bf{H}}_{b,k}}{\bf{F}}_{{\rm{RF,}}b}^{t + 1}{\bf{f}}_{{\rm{BB,}}b,k}^{t + 1}, \notag \\
& {\bf{g}}_b^t = {\left[ {g_{b,1}^t,\cdots ,g_{b,K}^t} \right]^T},  \;\;  {\Upsilon}_{b,k,j}^t = {\bf{w}}_k^H{{\bf{H}}_{b,k}}{\bf{F}}_{{\rm{RF,}}b}^{t + 1}{\bf{f}}_{{\rm{BB,}}b,j}^{t + 1}, \;\;  {\bf{L}}_b^t = {\rm{diag}}( {\hat L_{b,1}^t, \cdots ,\hat L_{b,K}^t}), \notag \\
& {\bf{l}}_b^t = {[ {\hat l_{b,1}^t, \cdots ,\hat l_{b,K}^t}]^T}, \;\; \hat l_{b,j}^t = \sum\limits_{k = 1}^K {{{\left| {{\xi _k}} \right|}^2}{\Upsilon}_{b,k,j}^{t,H}\hat C_{b,k,j}^t}, \;\; \hat L_{b,j}^t = \sum\limits_{k = 1}^K {{{\left| {{\xi _k}} \right|}^2}{\Upsilon}_{b,k,j}^{t,H}{\Upsilon}_{b,k,j}^t},
\end{align}
and the specific derivation process for problem (\ref{SecIVBS17}) is provided in Appendix B.

Problem (\ref{SecIVBS17}) is a classical BIQP problem with linear integer constraints (\ref{SecIVBS17a}) and (\ref{SecIVBS17b}). In detail, (\ref{SecIVBS17a}) stems from the fact that each user is served by at least one BS while (\ref{SecIVBS17b}) implies that $K_b$ users are served by $b$th BS concurrently. To this end, the RLA method is proposed to solve problem (\ref{SecIVBS17}), which replaces each quadratic term in the objective function by a new binary variable and two inequality constraints. Then, problem (\ref{SecIVBS17}) can be equivalently written as
\begin{align}\label{SecIVBS19}
& \mathop {\min }\limits_{{{\bm{\tau }}_b}} \; \sum\limits_{i = 1}^K {\sum\limits_{j = 1}^K {{u_{b,i,j}^t}{\tau _{b,i}}{\tau _{b,j}}} }  + \sum\limits_{i = 1}^K {{r_{b,i}^t}{\tau _{b,i}}} \notag \\
& \; {\rm{s.t.}} \;\; {{\bm{\tau }}_b} \in {{\cal G}_b},
\end{align}
where ${u_{b,i,j}^t}$ denotes $(i,j)$th entry of ${{\bf{U}}_b^t}$, ${r_{b,i}^t}$ denotes $i$th entry of ${{\bf{r }}_b^t}$, and
\begin{equation}\label{SecIVBS20}
\begin{aligned}
{{\cal G}_b} = \left\{ {{{\bm{\tau }}_b} \in {{\left\{ {0,1} \right\}}^K}:\sum\limits_{i \in {\cal K}} {{\tau _{b,i}} = {K_b},} {\rm{ }}{\tau _{b,i}} \ge {z_{b,i}},i \in {\cal K}} \right\}.
\end{aligned}
\end{equation}

To eliminate the quadratic terms in problem (\ref{SecIVBS19}), we use a new variable ${{\zeta}_{b,i,j}}$  to substitute for each product ${\tau _{b,i}}{\tau _{b,j}}$, $i,j \in {\cal K},i < j$. For the special case $i = j$, we can get ${{\zeta}_{b,i,i}} = {\tau _{b,i}}{\tau _{b,i}} = {\tau _{b,i}}$ due to the constraint ${\tau _{b,i}} \in \left\{ {0,1} \right\}$, $i \in {\cal K}$. By introducing ${{\zeta}_{b,i,j}} = {\tau _{b,i}}{\tau _{b,j}}$, the extra constraints can be expressed as
\begin{equation}\label{SecIVBS21}
\begin{aligned}
{{\zeta}_{b,i,j}}  = \max \left\{ {{\tau _{b,i}} + {\tau _{b,j}} - 1,0} \right\}, \;\;\; {{\zeta}_{b,i,j}}  = \min \left\{ {{\tau _{b,i}},{\tau _{b,j}}} \right\}, \;\; {{\zeta}_{b,i,j}} \in \left\{ {0,1} \right\}.
\end{aligned}
\end{equation}

Considering that  ${{\bf{U}}_b} \in {{\mathbb C}^{K \times K}}$ is symmetric and positive semidefinite, problem (\ref{SecIVBS19}) can be reformulated as
\begin{subequations}\label{SecIVBS22}
\begin{align}
&\mathop {\min }\limits_{{{\bm{\tau }}_b}} {\rm{ }}\sum\limits_{i = 1}^K {\sum\limits_{j = 1,j > i}^K {2{u_{b,i,j}^t}{{\zeta}_{b,i,j}}} }  + \sum\limits_{i = 1}^K {\left( {{u_{b,i,i}^t} + {r_{b,i}^t}} \right){\tau _{b,i}}} \label{SecIVBS22main} \\
& \; {\rm{s.t.}} \;\; {{\bm{\tau }}_b} \in {{\cal G}_b}, \label{SecIVBS22a}\\
& \;\;\;\;\;\;\;\; {{\zeta}_{b,i,j}} \le {\tau _{b,i}},i,j \in {\cal K},i < j, \label{SecIVBS22b} \\
& \;\;\;\;\;\;\;\; {{\zeta}_{b,i,j}} \le {\tau _{b,j}},i,j \in {\cal K},i < j, \label{SecIVBS22c} \\
& \;\;\;\;\;\;\;\; {{\zeta}_{b,i,j}} \ge {\tau _{b,i}} + {\tau _{b,j}} - 1,i,j \in {\cal K},i < j, \label{SecIVBS22d} \\
& \;\;\;\;\;\;\;\; {{\zeta}_{b,i,j}} \in \left\{ {0,1} \right\},i,j \in {\cal K},i < j. \label{SecIVBS22e}
\end{align}
\end{subequations}

The standard linearization process in problem (\ref{SecIVBS22}) neglects some intrinsic constraints \cite{PCF01}, so we propose an incremental RLA method that considers stronger constraints in the following. Specifically, given  ${\tau _{b,i}} = 1$ $(\forall i \in {\cal K})$, we can get ${{\zeta}_{b,i,j}} = {\tau _{b,j}}$ $(\forall j \in {\cal K})$, which further implies that $({{\zeta}_{b,i,1}},{{\zeta}_{b,i,2}}, \cdots ,{{\zeta}_{b,i,K}}) \in {{\cal G}_b}$ in the case of ${\tau _{b,i}} = 1$. Conversely, if ${\tau _{b,i}} = 0$, then ${{\zeta}_{b,i,j}} = 0$ $(\forall j \in {\cal K})$. Hence, the constraint related to ${\tau _{b,i}}$ $(\forall i \in {\cal K})$ can be represented as
\begin{align}\label{SecIVBS23}
{{\cal D}_b}\left( {{\tau _{b,i}}} \right) \buildrel \Delta \over = \left\{ {\begin{array}{*{20}{c}}
{{{\bf{0}}_K}, \; {\tau _{b,i}} = 0}\\
{{{\cal G}_b}, \; {\tau _{b,i}} = 1}
\end{array}} \right.
\end{align}
where ${{\cal D}_b}$ in the case of ${\tau _{b,i}} = 1$ can be given by
\begin{align}\label{SecIVBS24}
{{\cal D}_b}\left( {{\tau _{b,i}}} \right)= \left\{ {({{\zeta}_{b,i,1}}, \cdots ,{{\zeta}_{b,i,K}}) \in {{\left\{ {0,{\tau _{b,i}}} \right\}}^K}:} \sum\nolimits_{j \in {\cal K}} {{{\zeta}_{b,i,j}} = {K_b}{\tau _{b,i}}}, {{{\zeta}_{b,i,j}} \ge {z_{b,j}}{\tau _{b,i}}} \right\}.
\end{align}

Eventually, the BS selection problem for the P-CF-mMIMO system can be formulated as
\begin{subequations}\label{SecIVBS25}
\begin{align}
& \mathop {\min }\limits_{{{\bm{\tau }}_b}} {\rm{ }}\sum\limits_{i = 1}^K {\sum\limits_{j = 1,j > i}^K {2{u_{b,i,j}^t}{{\zeta}_{b,i,j}}} }  + \sum\limits_{i = 1}^K {\left( {{u_{b,i,i}^t} + {r_{b,i}^t}} \right){\tau _{b,i}}} \label{SecIVBS25main} \\
&\; {\rm{s.t.}} \;\; ({{\zeta}_{b,i,1}},{{\zeta}_{b,i,2}}, \cdots ,{{\zeta}_{b,i,K}}) \in {{\cal D}_b}\left( {{\tau _{b,i}}} \right), \; \forall i \in {\cal K}, \label{SecIVBS25a}\\
& \; \;\;\;\;\;\;\;{\text{(\ref{SecIVBS22a}), (\ref{SecIVBS22b}), (\ref{SecIVBS22c}), (\ref{SecIVBS22d}), (\ref{SecIVBS22e})}}. \label{SecIVBS25b}
\end{align}
\end{subequations}

As discussed above,  our proposed RLA algorithm transforms the BIQP problem into binary integer linear programming (BILP) problem. Remarkably, the BILP problem can be easily handled by existing general-purpose solvers \cite{PCF02}, including but not limited to branch-and-bound, cutting plane and Lagrangian duality.

\subsection{Fix \protect\boldmath$({\bf{\hat F}},{\bf{\hat \lambda }},{\bf{\hat \xi }})$  and Solve $({\bf{\hat w}}, {\bf{\hat \Phi }})$}
Due to the utilization of BS selection mechanism in the RIS-aided P-CF-mMIMO system, the BF optimization problems for solving ${\bf{\hat \Phi }}$ and ${\bf{\hat w}}$ also need to be reformulated. According to problem (\ref{SecIVBS06}), the combiner design at users can be formulated as
%
\begin{align}\label{SecIVBS28}
& \mathop {{\rm{min}}}\limits_{{\bf{\hat w}}} \; {{{\bf{\hat w}}}^H}{\bf{\hat P\hat w}} - \Re \left( {2{{{\bf{\hat w}}}^H}{\bf{\hat q}}} \right)\notag \\
&\;  {\rm{s.t.}} \;\; {\left| {{\bf{\hat w}}\left( i \right)} \right|^2} = 1,\forall {i_3} = 1,2, \cdots ,K{N_r}.
\end{align}
where
\begin{align}\label{SecIVBS29}
& {{\bf{\bar q}}_k} = \sum\limits_{b = 1}^B {\sqrt {{\mu _k}} \xi _k^H{{\bf{H}}_{b,k}}{{\bf{F}}_{{\rm{RF,}}b}}{{\bf{f}}_{{\rm{BB,}}b,k}}{\tau _{b,k}}}, \;\; {{\bf{\bar p}}_{k,j}} = \sum\limits_{b = 1}^B {{{\bf{H}}_{b,k}}{{\bf{F}}_{{\rm{RF,}}b}}{{\bf{f}}_{{\rm{BB,}}b,j}}{\tau _{b,j}}}, \notag \\
& {\bf{\hat q}} = {\left[ {{\bf{\bar q}}_1^T, \cdots ,{\bf{\bar q}}_K^T} \right]^T}, \; {{\bf{\bar P}}_k} = \sum\limits_{j = 1}^K {{{\left| {{\xi _k}} \right|}^2}{{{\bf{\bar p}}}_{k,j}}{\bf{\bar p}}_{k,j}^H}, \; {\bf{\hat P}} = {\rm{diag}}\left( {{{{\bf{\bar P}}}_1}, \cdots ,{{{\bf{\bar P}}}_K}} \right).
\end{align}

Problem (\ref{SecIVBS28}) is a QCQP problem with unit-modulus constraints and we employ the MO algorithm to solve such a constrained problem.
To optimize ${\bf{\hat \Phi }}$ in the P-CF-mMIMO system, we extract the related terms from problem (\ref{SecIVBS06}). By defining ${\bm{\hat \varphi }} = {\bf{\hat \Phi }}{{\bf{1}}_{RM}}$, the optimization problem for designing ${\bf{\hat \Phi }}$ at RISs can be expressed as
\begin{align}\label{SecIVBS30}
& \mathop {{\rm{min}}}\limits_{\bm{\varphi }} \; {{{\bm{\hat \varphi }}}^H}{\bf{\hat Z} \bm{\hat \varphi }} + {{{\bm{\hat \varphi }}}^H}{\bm{\hat \kappa }} + {{{\bm{\hat \kappa }}}^H}{\bm{\hat \varphi }} \notag \\
&\; {\rm{s.t.}} \;\; {\left| {{\bm{\hat \varphi }}\left( {{i_2}} \right)} \right|^2} \le 1,\forall {i_2} = 1,2, \cdots ,RM,
\end{align}
where
\begin{align}\label{SecIVBS31}
& {{\bf{\hat u}}_{b,k}} = \sqrt {{\mu _k}} \xi _k^H {\rm{diag}}\left( {{\bf{w}}_k^H{{\bf{V}}_k}} \right){{\bf{G}}_b}{{\bf{F}}_{{\rm{RF,}}b}}{{\bf{f}}_{{\rm{BB,}}b,k}}{\tau _{b,k}}, \; {{\bf{\hat v}}_{b,k,j}} = {\rm{diag}}\left( {{\bf{w}}_k^H{{\bf{V}}_k}} \right){{\bf{G}}_b}{{\bf{F}}_{{\rm{RF,}}b}}{{\bf{f}}_{{\rm{BB,}}b,j}}{\tau _{b,j}}, \notag \\
& {\bm{\hat \kappa }} = \sum\limits_{k = 1}^K {\sum\limits_{j = 1}^K {\sum\limits_{b = 1}^B {{{\left| {{\xi _k}} \right|}^2}{{\hat E}_{k,j}}{\bf{\hat v}}_{b,k,j}^ * }  - \sum\limits_{k = 1}^K {\sum\limits_{b = 1}^B {{\bf{\hat u}}_{b,k}^ * } } } }, \;\; {\hat E_{k,j}} = \sum\limits_{b = 1}^B {{\bf{w}}_k^H{{{\bf{\bar H}}}_{b,k}}{{\bf{F}}_{{\rm{RF,}}b}}{{\bf{f}}_{{\rm{BB,}}b,j}}{\tau _{b,j}}},\notag \\
& {\bf{\hat Z}} = \sum\limits_{k = 1}^K {\sum\limits_{j = 1}^K {{{\left| {{\xi _k}} \right|}^2}\left( {\sum\limits_{b = 1}^B {{\bf{\hat v}}_{b,k,j}^ * } } \right)\left( {\sum\limits_{b = 1}^B {{\bf{\hat v}}_{b,k,j}^T} } \right)} }.
\end{align}

It is worth noting that $\bf{\hat Z}$ is positive semidefinite and the constraint set is convex. Therefore, similar to problem (\ref{SecIIIHB40}), the convex problem (\ref{SecIVBS31}) can be well solved by the PDS method.

\subsection{Complexity and Convergence Analyses}
The overall computational complexity of the AO algorithm in the RIS-aided CF-mMIMO system stems from updating the variable set $\left( {{\bf{F}},{\bf{\Phi }},{\bf{w}},{\bm{\lambda }},{\bm{\xi }}} \right)$. Firstly, the complexity of updating auxiliary variables $\left({\bm{\lambda }},{\bm{\xi }} \right)$ is determined by (\ref{SecIIIHB01}). Secondly, the complexity of HBF design at BSs comes from the ADMM algorithm, e.g., (\ref{SecIIIHB09a})-(\ref{SecIIIHB09d}). Specifically, since ${\bf{F}}_b$, ${\bf{F}}_{{\rm{BB,}}b}$ and ${\bf{\Delta }}$ have close-form formulas, the complexity of updating $({\bf{F}}_b, {\bf{F}}_{{\rm{BB,}}b}, {\bf{\Delta }})$ can be directly calculated by (\ref{SecIIIHB13}), (\ref{SecIIIHB29}) and (\ref{SecIIIHB09d}), respectively. The complexity of updating ${\bf{F}}_{{\rm{RF,}}b}$ comes from the MO method as well as the combining design of {\bf{w}} at users, and the complexity of optimizing ${\bf{\Phi }}$ at RISs is caused by the PDS method. The extra complexity in the P-CF-mMIMO system is from the RLA method, which adopts the cutting plane to solve the BILP problem \cite{PCF03}. To sum up, the detailed complexity of CF-mMIMO and P-CF-mMIMO systems is shown in Table \ref{tab1}, where ${t_{\max }}$ and ${I_{\max }}$ denotes the maximum iteration number of ADMM and AO, respectively.

\begin{table}
\begin{center}
\caption{Complexity Analysis.}
\label{tab1}
\begin{tabular}{| c | c | c |}
\hline
\small{Schemes} & \small{Number of complex multiplications}\\
\hline
\small{CF-mMIMO} & \footnotesize{$\begin{array}{l}
{I_{\max }}\left[ {2B{K^2}\left( {{N_t}{N_r} + {N_t}{N_{{\rm{RF}}}} + {N_{{\rm{RF}}}}} \right) + {t_{\max }}B\left( {2K{N_t}{N_{{\rm{RF}}}} + K{N_t}} \right.} \right. + KN_t^3N_{{\rm{RF}}}^2\\
 + {K^2}N_t^2N_{{\rm{RF}}}^2 + {K^2}{N_t}{N_r} + N_t^2N_{{\rm{RF}}}^2 + {K^3}N_{{\rm{RF}}}^3 + {K^3}{N_t}N_{{\rm{RF}}}^2 + {K^2}{N_t}{N_{{\rm{RF}}}} + KN_{{\rm{RF}}}^2\\
 + K{N_t}{N_r} + K{N_t}{N_{{\rm{RF}}}}\left. { + K{N_t}{N_{{\rm{RF}}}}} \right) + \left( {2B{N_t}{N_r} + 2B{N_{{\rm{RF}}}}{N_r} + {K^2}N_r^2 + KN_r^2} \right)\\
\left. { + {K^2}{R^2}{M^2} + \left( {{K^2}B + KB} \right)\left( {{R^2}{M^2} + RM{N_t} + {N_t}{N_{{\rm{RF}}}}} \right)} \right]
\end{array}$}\\
\hline
\small{P-CF-mMIMO} & \footnotesize{$\begin{array}{l}
{I_{\max }}\left[ {\sum\nolimits_{b \in {\cal B}} {K_b^2\left( {{N_t}{N_r} + {N_t}{N_{{\rm{RF}}}} + {N_{{\rm{RF}}}}} \right)}  + {t_{\max }}\sum\nolimits_{b \in {\cal B}} {\left( {2{K_b}{N_t}{N_{{\rm{RF}}}} + {K_b}{N_t} + {K_b}N_t^3N_{{\rm{RF}}}^2} \right.} } \right.\\
 + K_b^2N_t^2N_{{\rm{RF}}}^2 + K_b^2{N_t}{N_r} + N_t^2N_{{\rm{RF}}}^2 + K_b^3N_{{\rm{RF}}}^3 + K_b^3{N_t}N_{{\rm{RF}}}^2 + K_b^2{N_t}{N_{{\rm{RF}}}} + {K_b}N_{{\rm{RF}}}^2\\
 + {K_b}{N_t}{N_r} + {K_b}{N_t}{N_{{\rm{RF}}}}\left. { + {K_b}{N_t}{N_{{\rm{RF}}}}{{K_b^2\left( {5{K_b} + 1} \right)} \mathord{\left/
 {\vphantom {{K_b^2\left( {5{K_b} + 1} \right)} 2}} \right.
 \kern-\nulldelimiterspace} 2}} \right) + 2B{N_t}{N_r} + 2B{N_{{\rm{RF}}}}{N_r}\\
 + K_b^2N_r^2 + {K_b}N_r^2\left. { + K_b^2{R^2}{M^2} + \left( {K_b^2B + KB} \right)\left( {{R^2}{M^2} + RM{N_t} + {N_t}{N_{{\rm{RF}}}}} \right)} \right]
\end{array}$} \\
\hline
\end{tabular}
\end{center}\vspace{-2em}
\end{table}

To prove the convergence of the AO algorithm in RIS-aided CF-mMIMO systems, we need to guarantee that each step of the iterative process is monotonous. Remarkably, the AO algorithm is achieved by alternately implementing ADMM, MO, and PDS methods, whose convergence properties have been demonstrated in \cite{System02}, \cite{System02b} and \cite{System03}, respectively. Besides, the convergence of the RLA method is proved in \cite{PCF02}. Considering that the WSR of practical CF-mMIMO and P-CF-mMIMO systems is finite, the AO algorithm is monotonous and eventually converges.

\section{Simulation Results}
In this section, simulation results are provided to demonstrate the effectiveness of proposed CBF algorithms in the RIS-aided CF-mMIMO network. Meanwhile, simulation results also reveals that, by using the BS selection strategy, the proposed P-CF-mMIMO network makes a better tradeoff between communication costs and WSR performance compared with the fully-connected CF-mMIMO network.

\subsection{Simulation Setup}
\begin{figure}[!t]
\centering
    \includegraphics[width=7cm]{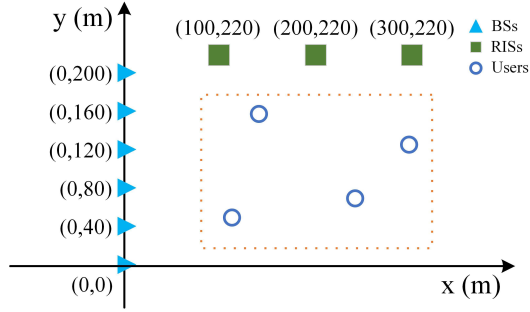}
\caption{Simulation setup of RIS-aided CF-mMIMO scenario.}\label{Simul00}\vspace{-2em}
\end{figure}

We consider a practical RIS-aided mmWave CF-mMIMO scenario as shown in Fig. \ref{Simul00}, where the physical positions of the $b$th BS and $r$th RIS are deployed at $\left( 0m, 40\times(b - 1)m \right)$ with the height of $6m$ and at $\left( (100 \times r)m, 220m \right)$ with the height of $4m$, respectively. Without loss of generality, all the users are independently and uniformly distributed within the $240m \times 160m$  rectangular area, and the height of each user is $1.8 m$. Given the 3D coordinate information, the LoS distance between any two points can be obtained. The number of array elements for each BS, each RIS and each user is set as $N_t=32$, $M=64$  and  $N_r=8$, respectively. Considering the sparse feature of mmWave channel, each communication link (e.g., ${{\bf{\bar H}}_{b,k}}$, ${{\bf{G}}_{b,r}}$ and ${{\bf{V}}_{r,k}}$) is composed of $L=4$ propagation paths, including $1$ LoS path and $3$ NLoS paths. More specifically, the complex gain of the LoS path is calculated by free space loss and the complex gains of the NLoS paths are calculated by multipath propagation loss \cite{Simulation01}. The AoAs and AoDs of the mmWave channels are selected from $(0,2\pi )$  following the uniform distribution. In addition, the working frequency of the RIS-aided CF-mMIMO system is set as $f_c=28$ GHz and the received noise power is ${\sigma ^2} =  - 85$ dBm.  For simplification, the maximum transmit power for all $B$ BSs is set as $P_b=P_{max}$, $\forall b \in {\cal B}$.

\subsection{Performance Comparisons}
%

\begin{figure}[!t]
\centering
\subfloat[]{
\begin{minipage}[t]{0.5\linewidth}
\centering
\includegraphics[width=6.5cm]{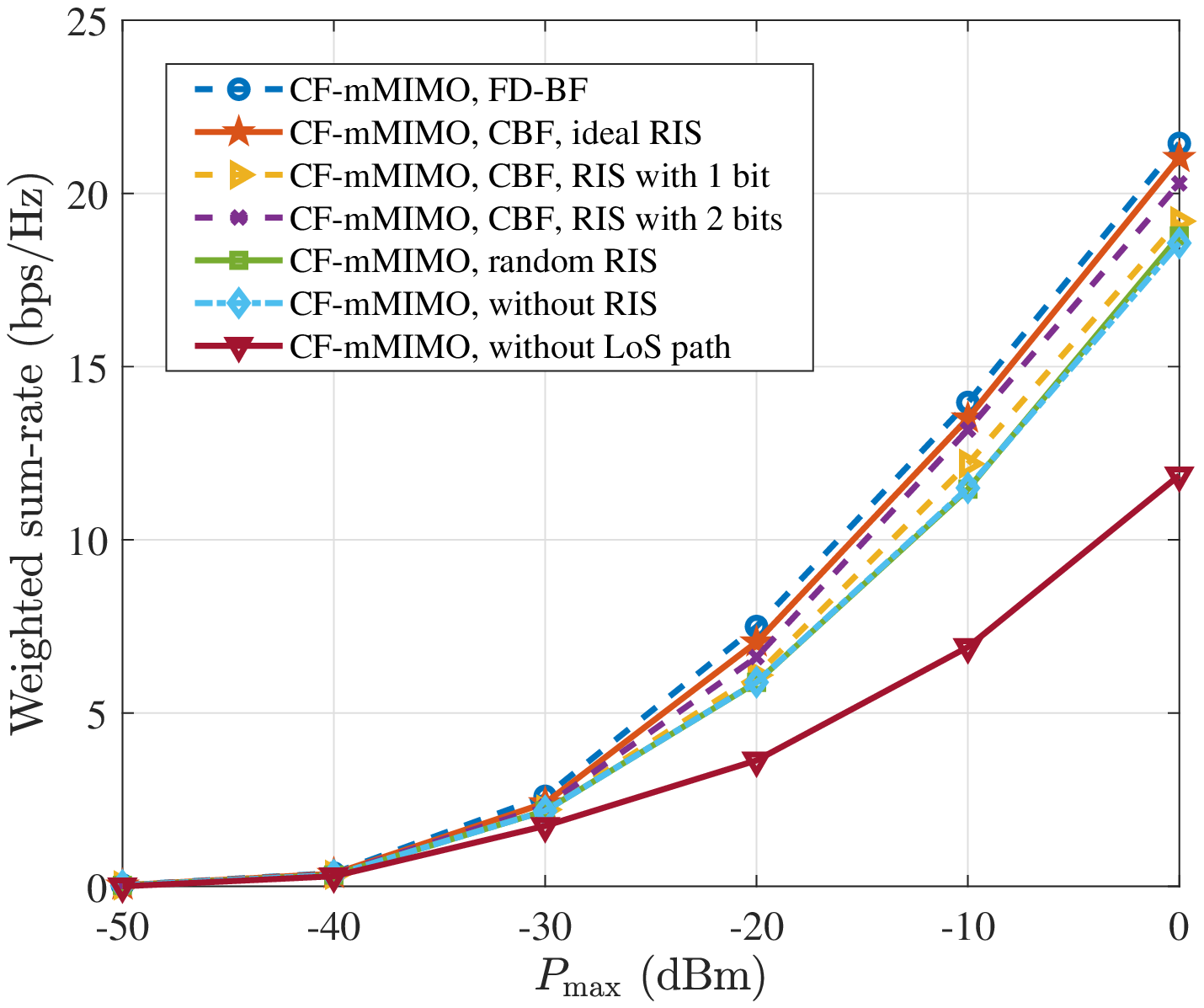}\label{Simul01}
\end{minipage}%
}
\subfloat[]{
\begin{minipage}[t]{0.5\linewidth}
\centering
\includegraphics[width=6.5cm]{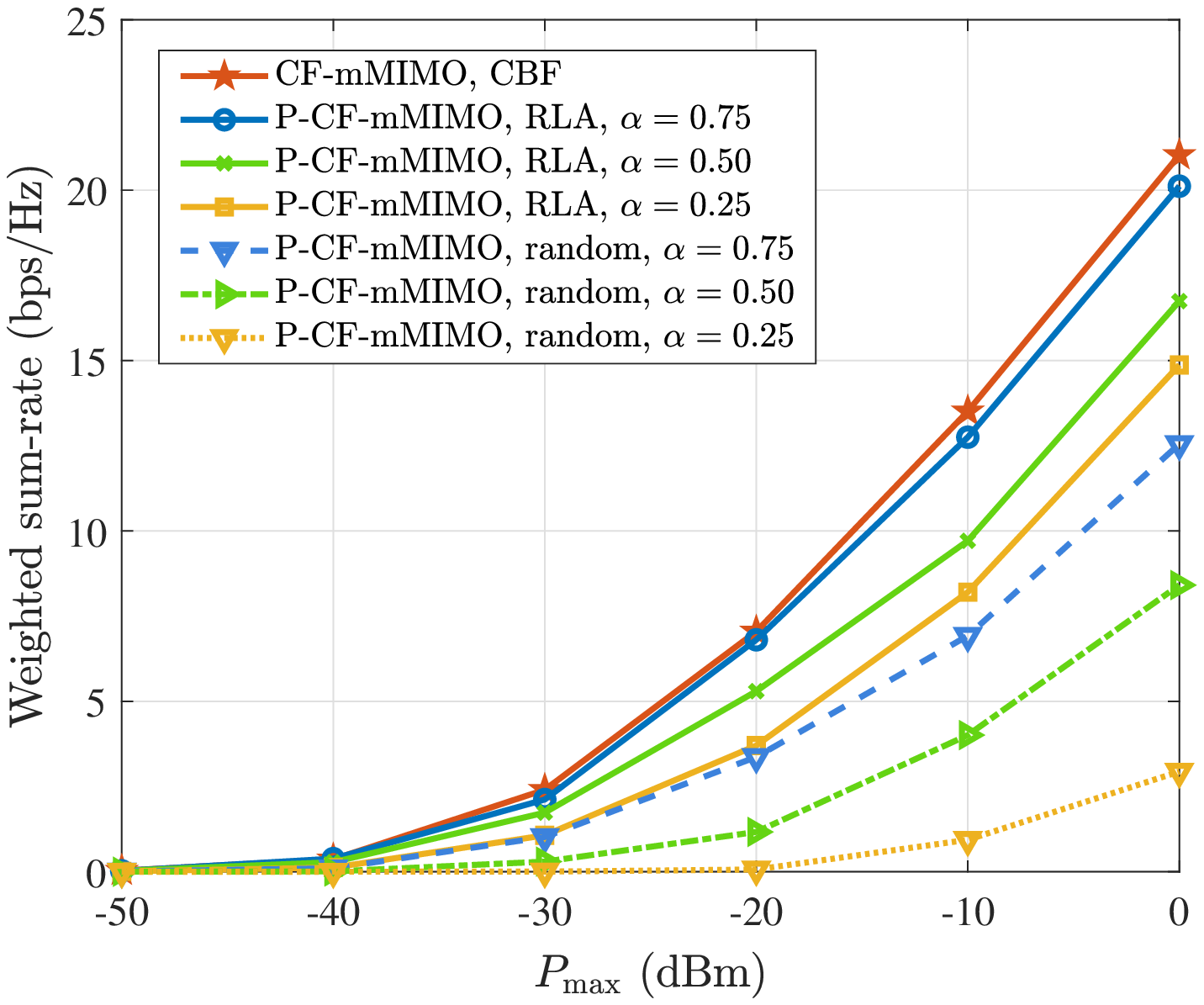}\label{Simul02}
\end{minipage}
}
\caption{\small{WSR versus transmit power $P_{max}$: (a) CF-mMIMO; (b) P-CF-mMIMO.}}\label{Simul012}\vspace{-2em}
\end{figure}

Fig. \ref{Simul012} investigates WSR comparisons versus the transmit power $P_{max}$ for both CF-mMIMO and P-CF-mMIMO systems, where $B=6$, $R=3$ and $K=4$. From Fig. \ref{Simul012}(a), it can be observed that the WSR of all the considered schemes grows rapidly with the increase of $P_{max}$ at BSs. Particularly, the WSR of our proposed CBF with ideal RIS approaches the fully-digital BF (FD-BF) that owns the optimal performance. With regard to the non-ideal RIS cases, the proposed CBF using RISs with $2$ bits achieves the similar WSR performance compared with the ideal RIS case, while deploying RISs with $1$ bit suffers from about $1.12$ bps/Hz performance penalty at $P=-10$ dBm. To validate the benefit of RISs, we also present the simulation curve without RISs that has worse WSR performance than RIS-aided schemes. Fig. \ref{Simul012}(a) also reveals that LoS path is essential for high frequency communications.
Fig. \ref{Simul012}(b) depicts the WSR performance of the RIS-aided P-CF-mMIMO system that can be controlled by adjusting the parameter ${\alpha }$. More precisely, as the value of ${\alpha }$ increases, the WSR performance of the RIS-aided P-CF-mMIMO system becomes better, and the communication costs grow accordingly. Under the same setting of ${\alpha }$, our proposed RLA based BS selection scheme greatly outperforms the random BS selection, indicating that the integer programming couples well with the BS selection problem. Remarkably, the RLA empowered RIS-aided P-CF-mMIMO system with ${\alpha }=0.75$ is capable of achieving basically the same WSR compared with the RIS-aided CF-mMIMO system (e.g., $0.64$ bps/Hz gap), which proves the effectiveness of our proposed RLA based BS selection scheme.

\begin{figure}[!t]
\centering
\subfloat[]{
\begin{minipage}[t]{0.5\linewidth}
\centering
\includegraphics[width=6.5cm]{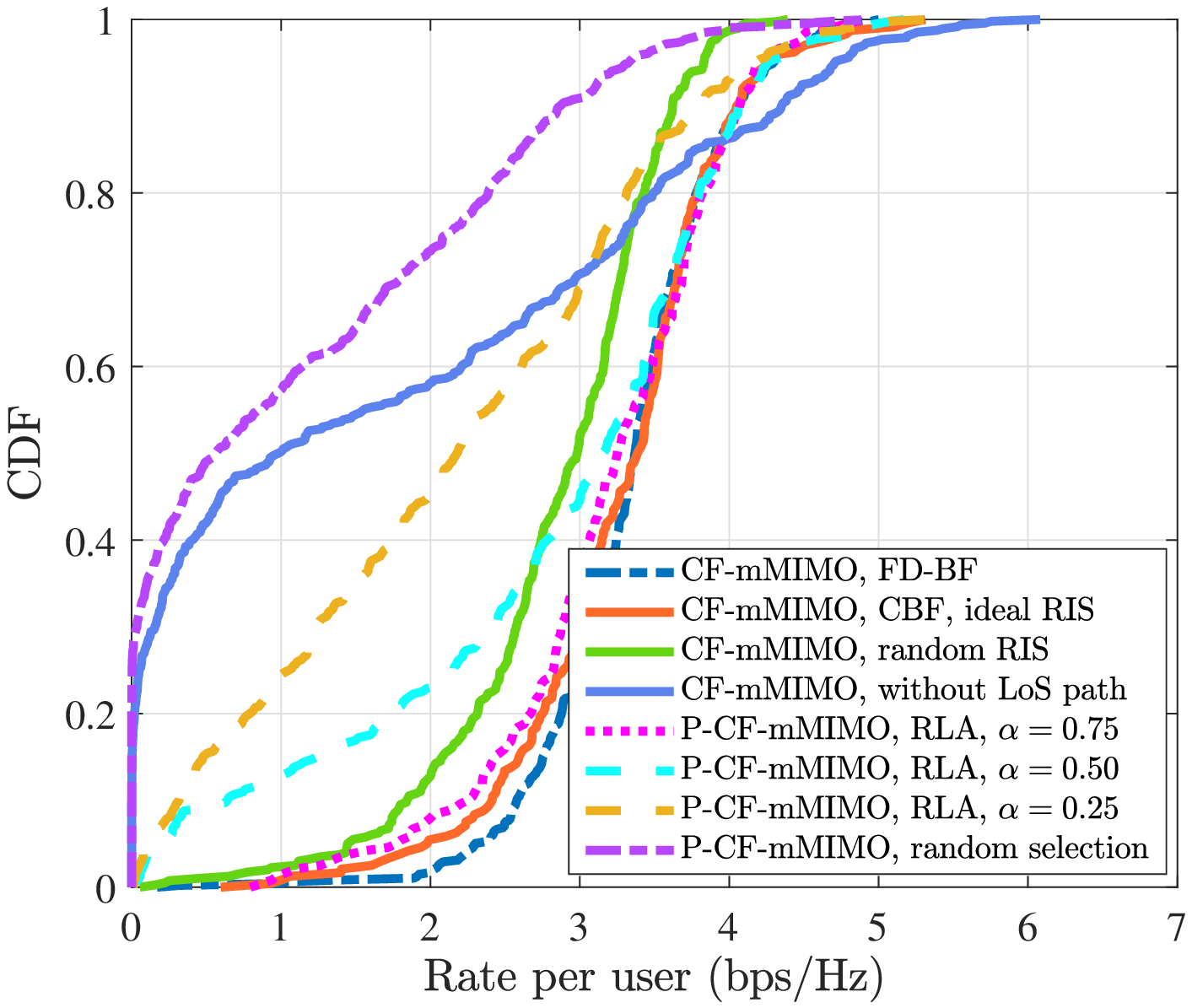}
\label{Simul03a}
\end{minipage}%
}
\subfloat[]{
\begin{minipage}[t]{0.5\linewidth}
\centering
\includegraphics[width=6.5cm]{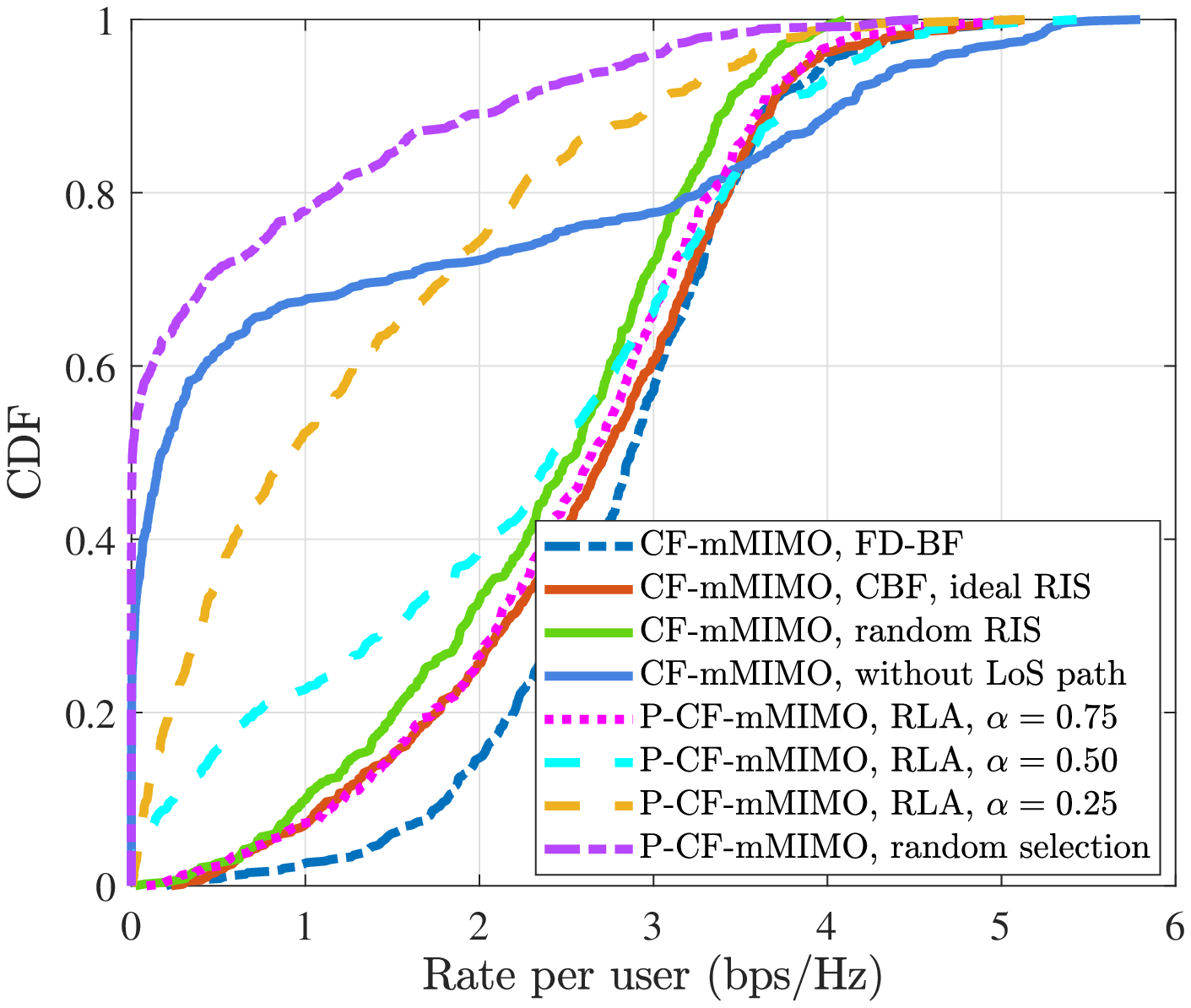}
\label{Simul03b}
\end{minipage}
}
\caption{CDF versus rate per user: (a) $K=4$; (b) $K=8$.}\label{Simul03}\vspace{-2em}
\end{figure}

Fig. \ref{Simul03} explores the cumulative distribution function (CDF) of rate per user with different user densities in both CF-mMIMO and P-CF-mMIMO systems, where all the users are randomly distributed in the predefined area. As a whole, Fig. \ref{Simul03}(a) and Fig. \ref{Simul03}(b) have a similar curve tendency. In Fig. \ref{Simul03}(a), when it comes to the CF-mMIMO system, the FD-BF scheme achieves the best per user rate performance, and outstrips the proposed CBF scheme with ideal RISs in terms of $95\%$-likely per user rate ($2.28$ bps/Hz versus $2.12$ bps/Hz) and basically the same median per user rate, implying that these performance gaps can be negligible. Moreover, the curves with random RISs in Fig. \ref{Simul03}(a) endure obvious performance loss compared to the case with ideal RISs, which indicates that RISs produce a dramatic impact on enhancing the network capacity. Another important observation in Fig. \ref{Simul03}(a) is that the CDF of the RLA empowered P-CF-mMIMO system with ${\alpha}=0.75$ is basically consistent with the CF-mMIMO system, where the performance gap of $95\%$-likely per user rate and median rate are $0.28$ bps/Hz and $0.13$ bps/Hz, respectively. The random BS selection scheme has the worst per user rate performance, and in turn verifies the effectiveness of our proposed RLA based BS selection scheme. Moreover, from Fig. \ref{Simul03}(a) and Fig. \ref{Simul03}(b), we can observe that smaller user density case with $K = 4$ owns much better per user rate than the large user density case with $K = 8$. The intuition is that more users certainly bring higher IUIs and worsen the per user rate.


%


\begin{figure}
\begin{minipage}{.49\textwidth}
\centering
\includegraphics[width = 6.5cm]{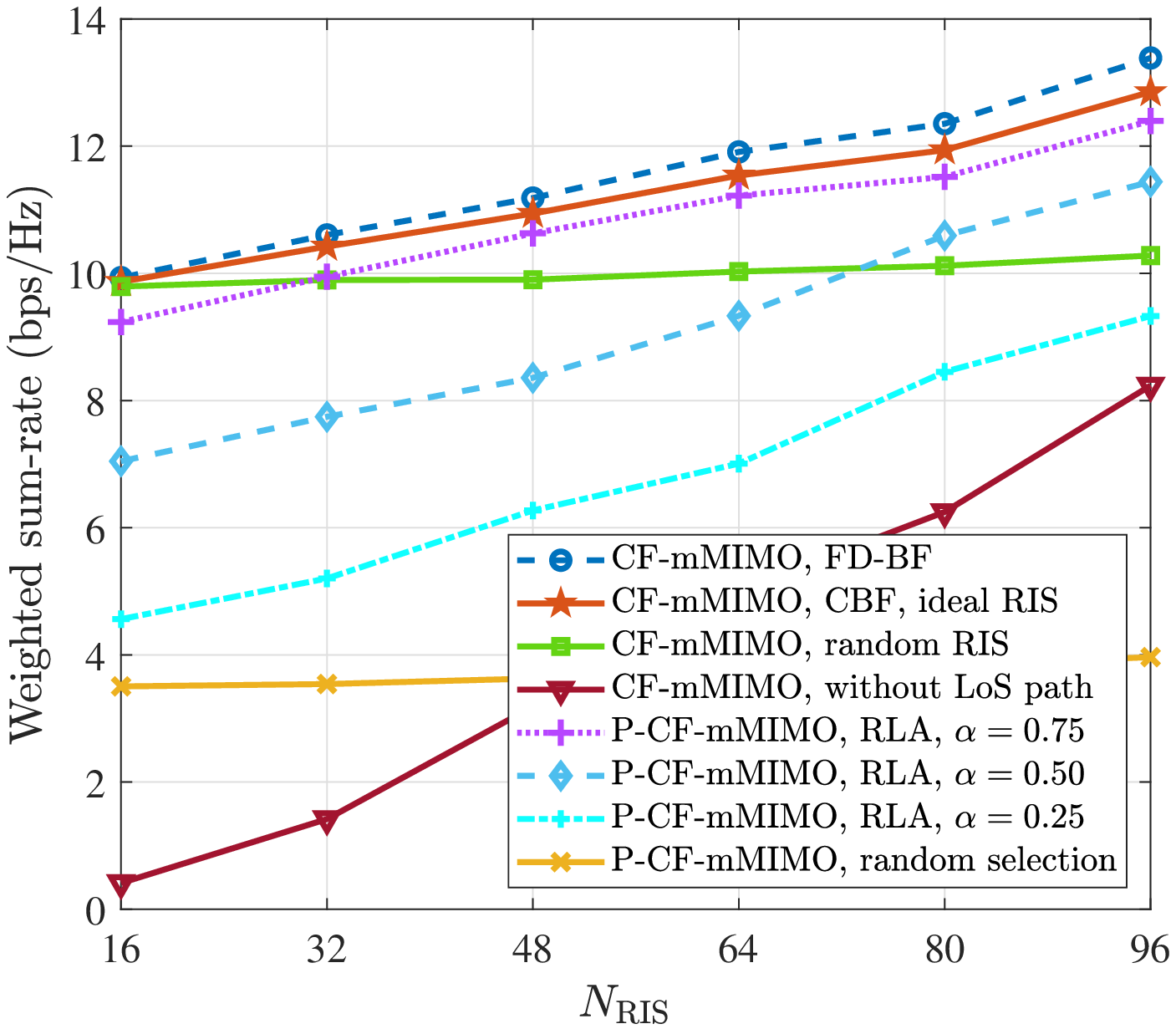}
\vskip-1.5ex
\caption{WSR versus the number of reflecting elements $N_{\text{RIS}}$.} \label{Simul04}
\end{minipage}
~\begin{minipage}{.49\textwidth}
\centering
\includegraphics[width = 6.5cm]{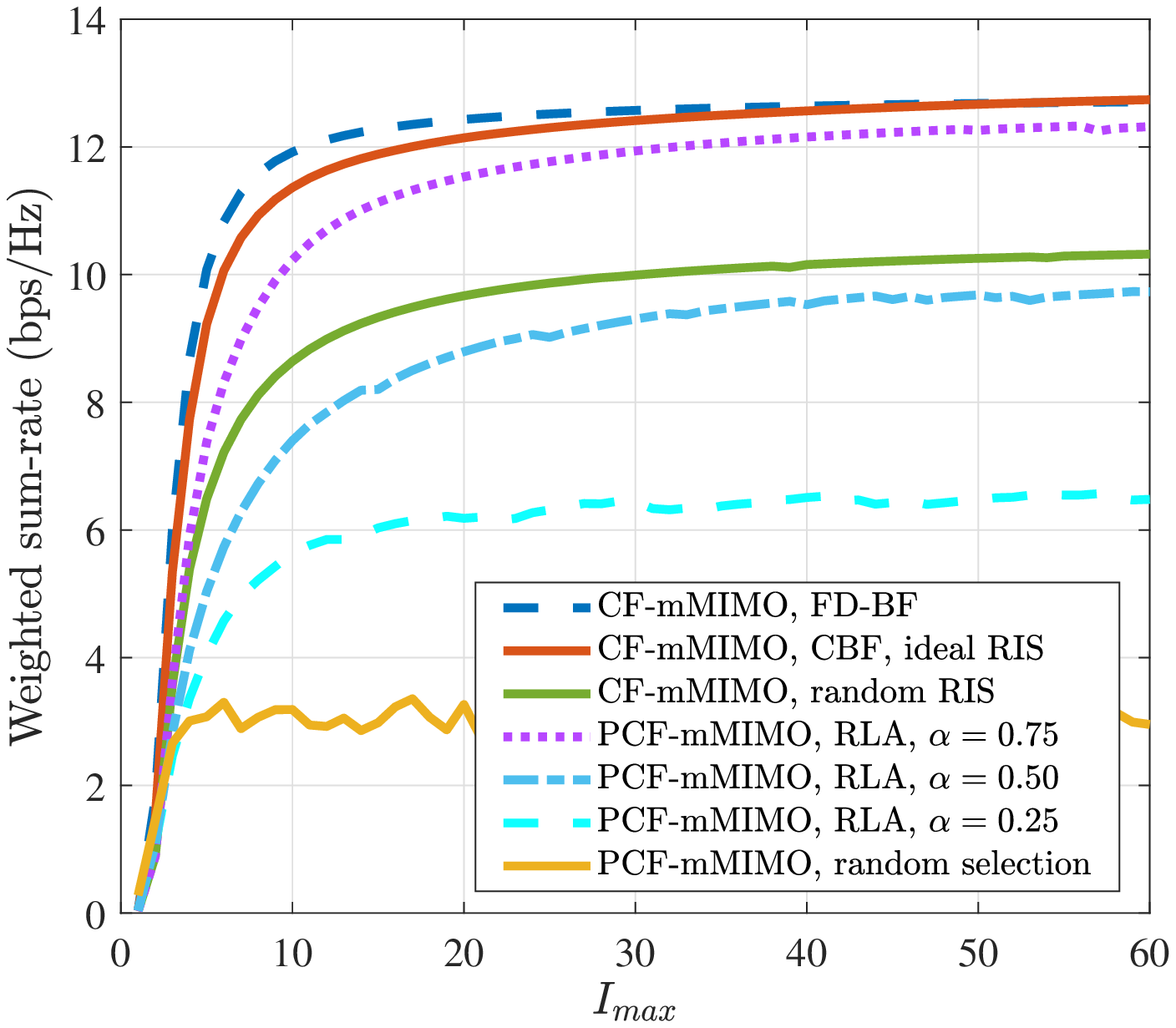}
\vskip-1.5ex
\caption{Sum-rate versus the number of iterations $I_{max}$.} \label{Simul08}
\end{minipage}
\vspace{-2em}
\end{figure}

Fig. \ref{Simul04} illustrates the sum-rate comparisons of considered schemes with the increasing number of RIS elements. Except for the CF-mMIMO system with random RISs and the P-CF-mMIMO system with random BS selection, the WSR of all the remaining schemes continuously grows as the number of reflecting elements rises. In particular, RISs have a greater impact on the CF-mMIMO system without LoS path. The reason is that the case without LoS path mainly depends on the BS-RIS channel and RIS-user channel to achieve high WSR performances, and thus RISs become more essential for the case without LoS path. For instance, the WSR gap between the CBF with ideal RISs and the case without LoS is around 8.18 bps/Hz at ${N_{{\rm{RIS}}}} = 32$  and 4.61 bps/Hz at ${N_{{\rm{RIS}}}} = 96$, respectively. Furhtermore, the performance gap between RLA based BS selection scheme and the random BS selection scheme grows larger with the increasing number of RIS elements, which also proves the significance of our proposed RLA based scheme for the RIS-aided P-CF-mMIMO system.

Fig. \ref{Simul08} presents the convergence performance of considered schemes versus the number of iterations in both CF-mMIMO and P-CF-mMIMO systems, where the parameter setup is $B=6$, $R=3$ and $K=4$. It can be easily observed that the WSR performance of all the considered schemes rises along with the increasing number of iterations, and then gradually slows down until convergence. Particularly, the converged number of iterations for both CF-mMIMO and P-CF-mMIMO systems is around ${I_{\max }} = 30$. Under the condition of ${I_{\max }} \ge 30$, the WSR of the RLA based P-CF-mMIMO system with ${\alpha}=0.75$ approaches the CBF scheme of the conventional CF-mMIMO system with ideal RISs (about 0.35 bps/Hz gap). Nevertheless, at the beginning of the iterative process, the RLA based scheme converges slower than the FD-BF and proposed CBF schemes. This phenomenon stems from that the P-CF-mMIMO system has to optimize the extra BS selection matrix ${\bf{{\Lambda}}}_b$ and thus results in slower network convergence rate. In addition, compared with random BS selection scheme, the RLA based BS selection scheme reaps obvious WSR improvement. By choosing different $\alpha$, our proposed RLA based P-CF-mMIMO system is able to balance communication costs and WSR performances.

\section{Conclusion}
This paper firstly investigated the active and passive BF design for the RIS-aided mmWave CF-mMIMO system. To decrease communication costs brought by fully-connected CF-mMIMO systems, we proposed a novel P-CF-mMIMO framwork that a limited number of communication links were connected among users and BSs. The BS selection problem was modeled as a BIQP problem and then we developed an incremental RLA scheme to solve this integer programming problem. By selecting appropriate NCR, our proposed P-CF-mMIMO systemd achieved a better compromise between performance and communication costs, which is able to meet diverse requirements in future 6G wireless networks.

\begin{appendices}
\section{The Detailed Proof for Problem (\ref{SecIIIHB16})}
From problem (\ref{SecIIIHB14}) we note that the objective function is composed of three parts. To obtain problem (\ref{SecIIIHB16}), the first term of ${f_2}( {{{\bf{F}}_{{\rm{RF}},b}}})$  can be reformulated as
\begin{align}\label{Appendix01}
& \left\| {{\bf{F}}_b^{t + 1} - {{\bf{F}}_{{\rm{RF,}}b}}{\bf{F}}_{{\rm{BB,}}b}^t + \frac{{{{\bf{\Delta }}^t}}}{{{\rho _1}}}} \right\|_F^2  = \left\| {{\rm{vec}}\left( {{\bf{F}}_b^{t + 1} - {{\bf{F}}_{{\rm{RF,}}b}}{\bf{F}}_{{\rm{BB,}}b}^t + \frac{{{{\bf{\Delta }}^t}}}{{{\rho _1}}}} \right)} \right\|^2 = \left\| {{\bf{B}}_b^t{{\bf{f}}_{{\rm{RF}},b}} - {\bf{m}}_b^t} \right\|^2 \notag \\
& = {\bf{f}}_{{\rm{RF}},b}^H{\bf{B}}_b^{t,H}{\bf{B}}_b^t{{\bf{f}}_{{\rm{RF}},b}} - {\bf{m}}_b^{t,H}{\bf{B}}_b^t{{\bf{f}}_{{\rm{RF}},b}} - {\bf{f}}_{{\rm{RF}},b}^H{\bf{B}}_b^{t,H}{\bf{m}}_b^t + {{\bf{m}}_b^{t,H}{\bf{m}}_b^t}.
\end{align}

Then, the second term of ${f_2}( {{{\bf{F}}_{{\rm{RF}},b}}})$ can be reformulated as
\begin{align}\label{Appendix02}
& - 2\Re \left( {\sum\limits_{k = 1}^K {\sqrt {{\mu _k}} \xi _k^H{\bf{w}}_k^{t,H}{{\bf{H}}_{b,k}}{{\bf{F}}_{{\rm{RF,}}b}}{\bf{f}}_{{\rm{BB,}}b,k}^t} } \right) =  - 2\Re \left( {\sum\limits_{k = 1}^K {\sqrt {{\mu _k}} \xi _k^H {\rm{vec}}\left( {{\bf{w}}_k^{t,H}{{\bf{H}}_{b,k}}{{\bf{F}}_{{\rm{RF,}}b}}{\bf{f}}_{{\rm{BB,}}b,k}^t} \right)} } \right) \notag \\
& =  - 2\Re \left( {\sum\limits_{k = 1}^K {\sqrt {{\mu _k}} \xi _k^H{\bf{f}}_{{\rm{BB,}}b,k}^{t,T} \otimes \left( {{\bf{w}}_k^H{{\bf{H}}_{b,k}}} \right){{\bf{f}}_{{\rm{RF}},b}}} } \right) =  - {\bf{b}}_b^{t,H}{{\bf{f}}_{{\rm{RF}},b}} - {\bf{f}}_{{\rm{RF}},b}^H{\bf{b}}_b^t.
\end{align}

Subsequently, the last term of ${f_2}( {{{\bf{F}}_{{\rm{RF}},b}}})$ can be reformulated as
\begin{align}\label{Appendix03}
& \sum\limits_{k = 1}^K {\sum\limits_{j = 1}^K {{{\left| {{\xi _k}} \right|}^2}{{\left| {{\bf{w}}_k^{t,H}{{\bf{H}}_{b,k}}{{\bf{F}}_{{\rm{RF,}}b}}{\bf{f}}_{{\rm{BB,}}b,j}^t + C_{ b,k,j}^t} \right|}^2}} }
= \sum\limits_{k = 1}^K {\sum\limits_{j = 1}^K {{{\left| {{\xi _k}} \right|}^2}{{\left| {{\rm{vec}}\left( {{\bf{w}}_k^{t,H}{{\bf{H}}_{b,k}}{{\bf{F}}_{{\rm{RF,}}b}}{\bf{f}}_{{\rm{BB,}}b,j}^t} \right) + C_{b,k,j}^t} \right|}^2}} } \notag \\
& = \sum\limits_{k = 1}^K {\sum\limits_{j = 1}^K {{{\left| {{\xi _k}} \right|}^2}{{\left| {{\bf{f}}_{{\rm{BB,}}b,j}^{t,T} \otimes \left( {{\bf{w}}_k^H{{\bf{H}}_{b,k}}} \right){{\bf{f}}_{{\rm{RF}},b}} + C_{ b,k,j}^t} \right|}^2}} } = \sum\limits_{k = 1}^K {\sum\limits_{j = 1}^K {{{\left| {{\xi _k}} \right|}^2}{{\left| {{\bf{a}}_{b,k,j}^t{{\bf{f}}_{{\rm{RF}},b}} + C_{ b,k,j}^t} \right|}^2}} } \notag  \\
& = {\bf{f}}_{{\rm{RF}},b}^H\left( {\sum\limits_{k = 1}^K {\sum\limits_{j = 1}^K {{{\left| {{\xi _k}} \right|}^2}{\bf{a}}_{b,k,j}^{t,H}{\bf{a}}_{b,k,j}^t} } } \right){{\bf{f}}_{{\rm{RF}},b}} + {\bf{f}}_{{\rm{RF}},b}^H\left( {\sum\limits_{k = 1}^K {\sum\limits_{j = 1}^K {{{\left| {{\xi _k}} \right|}^2}{\bf{a}}_{b,k,j}^{t,H}C_{ b,k,j}^t} } } \right) \notag \\
& + \left( {\sum\limits_{k = 1}^K {\sum\limits_{j = 1}^K {{{\left| {{\xi _k}} \right|}^2}C_{ b,k,j}^{t,H}{\bf{a}}_{b,k,j}^t} } } \right){{\bf{f}}_{{\rm{RF}},b}} + \sum\limits_{k =1}^K {\sum\limits_{j =1}^K {{{\left| {{\xi _k}} \right|}^2}C_{ b,k,j}^{t,H}C_{ b,k,j}^t} }.
\end{align}

By considering (\ref{Appendix01}), (\ref{Appendix02}) and (\ref{Appendix03}), the proof for problem (\ref{SecIIIHB16}) is completed.

\section{The Detailed Proof for Problem (\ref{SecIVBS17})}
The main challenge is to convert $ {{\hat f}_2}( {{{\bf{{\Lambda}}}_b}} )$ in to a solvable form. Hence, the first part of $ {{\hat f}_2}( {{{\bf{{\Lambda}}}_b}} )$ can be reformulated as
\begin{align}\label{Appendix04}
&{\left\| {{\bf{F}}_b^{t + 1} - {\bf{F}}_{{\rm{RF,}}b}^{t + 1}{\bf{F}}_{{\rm{BB,}}b}^{t + 1}{{\bf{{\Lambda}}}_b} + \frac{{{{\bf{\Delta }}^t}}}{{{\rho _2}}}} \right\|^2} = {\left\| {{\rm{vec}}\left( {{\bf{F}}_b^{t + 1} - {\bf{F}}_{{\rm{RF,}}b}^{t + 1}{\bf{F}}_{{\rm{BB,}}b}^{t + 1}{{\bf{{\Lambda}}}_b} + \frac{{{{\bf{\Delta }}^t}}}{{{\rho _2}}}} \right)} \right\|^2} \notag \\
& = {\left\| {{\bf{m}}_b^t - {\bf{M}}_b^t{\rm{vec}}\left( {{{\bf{{\Lambda}}}_b}} \right)} \right\|^2} = {\left\| {{\bf{m}}_b^t - {\bf{\tilde M}}_b^t{{\bm{\tau }}_b}} \right\|^2}.
\end{align}

In general, integer programming requires that all the involved parameters are real-valued, so the trick is to convert the complex-valued problem into real-valued case, which can be given by
\begin{align}\label{Appendix05}
{\bf{\hat m}}_b^t = \left[ {\begin{array}{*{20}{c}}
{\Re \left( {{\bf{m}}_b^t} \right)}\\
{\Im \left( {{\bf{m}}_b^t} \right)}
\end{array}} \right],{\bf{\hat M}}_b^t = \left[ {\begin{array}{*{20}{c}}
{\Re \left( {{\bf{\tilde M}}_b^t} \right)}\\
{\Im \left( {{\bf{\tilde M}}_b^t} \right)}
\end{array}} \right],
\end{align}

Considering (\ref{Appendix05}), we can rewrite (\ref{Appendix04}) as a real-valued form, given by
\begin{align}\label{Appendix06}
& {\left\| {{\bf{m}}_b^t - {{{\bf{\tilde M}}}_b}{{\bm{\tau }}_b}} \right\|^2}
 = {\left\| {\begin{array}{*{20}{c}}
{\Re \left( {{\bf{m}}_b^t} \right) - \Re \left( {{{{\bf{\tilde M}}}_b}{{\bm{\tau }}_b}} \right)} \\
{\Im \left( {{\bf{m}}_b^t} \right) - \Im \left( {{{{\bf{\tilde M}}}_b}{{\bm{\tau }}_b}} \right)}
\end{array}} \right\|^2}
= {\left\| {\left[ {\begin{array}{*{20}{c}}
{\Re \left( {{\bf{m}}_b^t} \right)}\\
{\Im \left( {{\bf{m}}_b^t} \right)}
\end{array}} \right] - \left[ {\begin{array}{*{20}{c}}
{\Re \left( {{{{\bf{\tilde M}}}_b}} \right)}\\
{\Im \left( {{{{\bf{\tilde M}}}_b}} \right)}
\end{array}} \right]{{\bm{\tau }}_b}} \right\|^2} \notag \\
& = {\left\| {{\bf{\hat m}}_b^t - {\bf{\hat M}}_b^t{{\bm{\tau }}_b}} \right\|^2}
= {{\bm{\tau }}_b^T{\bf{\hat M}}_b^{t,T}{\bf{\hat M}}_b^t{{\bm{\tau }}_b} - {\bf{\hat m}}_b^{t,T}{\bf{\hat M}}_b^t{{\bm{\tau }}_b}} - {\bm{\tau }}_b^T{\bf{\hat M}}_b^{t,T}{\bf{\hat m}}_b^t + {{\bf{\hat m}}_b^{t,T}{\bf{\hat m}}_b^t}.
\end{align}

Next, the second part of $ {{\hat f}_2}( {{{\bf{{\Lambda}}}_b}} )$ can be reformulated as
\begin{align}\label{Appendix07}
- 2\Re \left( {\sum\limits_{k = 1}^K {\sqrt {{\mu _k}} \xi _k^H{\bf{w}}_k^H{{\bf{H}}_{b,k}}{\bf{F}}_{{\rm{RF,}}b}^{t + 1}{\bf{f}}_{{\rm{BB,}}b,k}^{t + 1}{\tau _{b,k}}} } \right)
=  - 2\Re \left( {\sum\limits_{k = 1}^K {g_{b,k}^t{\tau _{b,k}}} } \right)
=  - 2\Re \left( {{\bf{g}}_b^{t,T}{{\bm{\tau }}_b}} \right).
\end{align}

To handle the quadratic term, the last part in ${{\hat f}_2}( {{{\bf{{\Lambda}}}_b}} )$ can be rewritten as
\begin{align}\label{Appendix08}
& \; \sum\limits_{k = 1}^K {\sum\limits_{j = 1}^K {{{\left| {{\xi _k}} \right|}^2}{{\left| {{\bf{w}}_k^H{{\bf{H}}_{b,k}}{\bf{F}}_{{\rm{RF,}}b}^{t + 1}{\bf{f}}_{{\rm{BB,}}b,j}^{t + 1}{\tau _{b,j}} + \hat C_{ b,k,j}^t} \right|}^2}} }
= \sum\limits_{k = 1}^K {\sum\limits_{j = 1}^K {{{\left| {{\xi _k}} \right|}^2}{{\left| {{\Upsilon}_{b,k,j}^t{\tau _{b,j}} + \hat C_{b,k,j}^t} \right|}^2}} } \notag \\
& = \sum\limits_{k = 1}^K {\sum\limits_{j = 1}^K {{{\left| {{\xi _k}} \right|}^2}{{\left( {{\Upsilon}_{b,k,j}^t{\tau _{b,j}} + \hat C_{b,k,j}^t} \right)}^H}\left( {{\Upsilon}_{b,k,j}^t{\tau _{b,j}} + \hat C_{b,k,j}^t} \right)} } \notag \\
& = \sum\limits_{j = 1}^K {\tau _{b,j}^H\hat L_{b,j}^t} {\tau _{b,j}} + \sum\limits_{j = 1}^K {\tau _{b,j}^H\hat l_{b,j}^t}
+ \sum\limits_{j = 1}^K {{\tau _{b,j}}\hat l_{b,j}^{t,H}}  + \sum\limits_{k = 1}^K {\sum\limits_{j = 1}^K {{{\left| {{\xi _k}} \right|}^2}\hat C_{b,k,j}^{t,H}\hat C_{b,k,j}^t} } \notag \\
& = {\bm{\tau }}_b^H{\bf{L}}_b^t{{\bm{\tau }}_b} + {\bm{\tau }}_b^H{\bf{l}}_b^t + {\bf{l}}_b^{t,H}{{\bm{\tau }}_b} + {\sum\limits_{k = 1}^K {\sum\limits_{j = 1}^K {{{\left| {{\xi _k}} \right|}^2}\hat C_{b,k,j}^{t,H}\hat C_{b,k,j}^t} }}.
\end{align}

By merging congeners of (\ref{Appendix06}), (\ref{Appendix07}) and (\ref{Appendix08}), we accomplish the proof for problem (\ref{SecIVBS17}).

\end{appendices}


\begin{thebibliography}{1}

\bibitem{introduction01}
P. Yang, Y. Xiao, M. Xiao \emph{et al.}, ``6G Wireless Communications: Vision and Potential Techniques,'' \emph{IEEE Network}, vol. 33, no. 4, pp. 70-75, Aug. 2019.


\bibitem{introduction02}
H. Q. Ngo, A. Ashikhmin, H. Yang \emph{et al.}, ``Cell-free massive mimo versus small cells,'' \emph{IEEE Trans. Wirel. Commun.}, vol. 16, no. 3, pp. 1834-1850, 2017.

\bibitem{introduction02a}
E. Nayebi, A. Ashikhmin, T. L. Marzetta \emph{et al.}, ``Precoding and Power Optimization in Cell-Free Massive MIMO Systems,'' \emph{IEEE Trans. Wirel. Commun.}, vol. 16, no. 7, pp. 4445-4459, July 2017.

\bibitem{introduction02b}
H. A. Ammar, R. Adve, S. Shahbazpanahi \emph{et al.}, ``User-Centric Cell-Free Massive MIMO Networks: A Survey of Opportunities, Challenges and Solutions,'' \emph{IEEE Commun. Surv. Tutor.}, vol. 24, no. 1, pp. 611-652, 2022.

\bibitem{introduction03}
T. C. Mai, H. Q. Ngo, and T. Q. Duong, ``Downlink spectral efficiency of cell-free massive MIMO systems with multi-antenna users,'' \emph{IEEE Trans. Commun.}, vol. 68, no. 8, pp. 4803-4815, Aug. 2020.

\bibitem{introduction04}
J. Zhang, J. Zhang, D. W. K. Ng \emph{et al.}, ``Improving Sum-Rate of Cell-Free Massive MIMO With Expanded Compute-and-Forward,'' \emph{IEEE Trans. Signal Process.}, vol. 70, pp. 202-215, 2022.

\bibitem{introduction05}
E. Bjornson and L. Sanguinetti, ``Making cell-free massive MIMO competitive with MMSE processing and centralized implementation,'' \emph{IEEE Trans. Wirel. Commun.}, vol. 19, no. 1, 2019, pp. 77-90.

\bibitem{introduction06a}
M. Xiao, S. Mumtaz, Y. Huang \emph{et al.}, ``Millimeter Wave Communications for Future Mobile Networks,'' \emph{IEEE J. Sel. Areas Commun.}, vol. 35, no. 9, pp. 1909-1935, Sept. 2017.

\bibitem{introduction06}
Z. Chen, X. Y. Ma, B. Zhang \emph{et al.}, ``A survey on terahertz communications,'' \emph{China Commun.}, vol. 16, no. 2, pp. 1-35, Feb. 2019.

\bibitem{introduction07}
C. Huang, A. Zappone, G. C. Alexandropoulos \emph{et al.}, ``Reconfigurable intelligent surfaces for energy efficiency in wireless communication,'' \emph{IEEE Trans. Wirel. Commun.}, vol. 18, no. 8, pp. 4157-4170, Aug. 2019.

\bibitem{introduction08}
X. Ma, Z. Chen, W. Chen \emph{et al.}, ``Intelligent Reflecting Surface Enhanced Indoor Terahertz Communication Systems,'' \emph{Nano Commun. Netw.}, vol. 24, pp. 100284, May. 2020.

\bibitem{introduction09}
Q. Wu and R. Zhang, ``Towards smart and reconfigurable environment: Intelligent reflecting surface aided wireless network,'' \emph{IEEE Commun. Mag.}, vol. 58, no. 1, pp. 106-112, Jan. 2020.

\bibitem{introduction09a}
C. Pan, H. Ren, K. Wang \emph{et al.}, ``Multicell MIMO communications relying on intelligent reflecting surfaces,'' \emph{IEEE Trans. Wirel. Commun.}, vol. 19, no. 8, pp. 5218-5233, Aug. 2020.

\bibitem{introduction09b}
L. Dai, B. Wang, M. Wang \emph{et al.}, ``Reconfigurable intelligent surface-based wireless communications: Antenna design, prototyping, and experimental results,'' \emph{IEEE Access}, vol. 8, pp. 45913-45923, Mar. 2020.

\bibitem{introduction09c}
C. Huang, S. Hu, G. C. Alexandropoulos, \emph{et al.}, ``Holographic MIMO surfaces for 6G wireless networks: opportunities, challenges, and trends'', \emph{IEEE Wirel. Commun.}, vol. 27, no. 5, pp. 118-125, Oct. 2020.

\bibitem{introduction09d}
E. Shi, J. Zhang, S. Chen \emph{et al.}, ``Wireless Energy Transfer in RIS-Aided Cell-Free Massive MIMO Systems: Opportunities and Challenges,'' \emph{IEEE Commun. Mag.}, vol. 60, no. 3, pp. 26-32, Mar. 2022,

\bibitem{introduction10}
B. Al-Nahhas, M. Obeed, A. Chaaban and M. J. Hossain, ``RIS-Aided Cell-Free Massive MIMO: Performance Analysis and Competitiveness,'' \emph{in Proc. IEEE Int. Conf. Commun. Workshops (ICC Workshops)}, 2021, pp. 1-6.

\bibitem{introduction11}
Q. N. Le, V. D. Nguyen, O. A. Dobre \emph{et al.}, ``Energy Efficiency Maximization in RIS-Aided Cell-Free Network With Limited Backhaul,'' \emph{IEEE Commun. Lett.}, vol. 25, no. 6, pp. 1974-1978, June 2021.

\bibitem{introduction12}
S. Huang, Y. Ye, M. Xiao \emph{et al.}, ``Decentralized Beamforming Design for Intelligent Reflecting Surface-Enhanced Cell-Free Networks,'' \emph{IEEE Wirel. Commun. Lett.}, vol. 10, no. 3, pp. 673-677, Mar. 2021.

\bibitem{introduction13}
Z. Zhang and L. Dai, ``A Joint Precoding Framework for Wideband Reconfigurable Intelligent Surface-Aided Cell-Free Network,'' \emph{IEEE Trans. Signal Process.}, vol. 69, pp. 4085-4101, 2021.

\bibitem{introduction14}
O. Demir, E. Bjornson and L. Sanguinetti, ``Foundations of User-Centric Cell-Free Massive MIMO,'' \emph{Found. Trends Signal Process.}, vol. 14, nos. 3-4, pp 162-472, 2021.

\bibitem{introduction15}
O. Bursalioglu, G. Caire, R. Mungara \emph{et al.}, ``Fog massive MIMO: A user-centric seamless hotspot architecture,'' \emph{IEEE Trans. Wirel. Commun.}, vol. 18, no. 1, pp. 559-574, Jan. 2019.

\bibitem{introduction16}
S. Buzzi and C. D Andrea, ``Cell-free massive MIMO: User-centric approach,'' \emph{IEEE Wirel. Commun. Lett.}, vol. 6, no. 6, pp. 706-709, Dec. 2017.

\bibitem{introduction17}
E. Bjornson and L. Sanguinetti, ``Scalable Cell-Free Massive MIMO Systems,'' \emph{IEEE Trans. Commun.}, vol. 68, no. 7, pp. 4247-4261, July 2020.

\bibitem{introduction18}
S. Buzzi, C. D Andrea, A. Zappone \emph{et al.}, ``User-Centric 5G Cellular Networks: Resource Allocation and Comparison With the Cell-Free Massive MIMO Approach,'' \emph{IEEE Trans. Wirel. Commun.}, vol. 19, no. 2, pp. 1250-1264, Feb. 2020.

\bibitem{introduction18a}
M. Alonzo, S. Buzzi, A. Zappone \emph{et al.}, ``Energy-Efficient Power Control in Cell-Free and User-Centric Massive MIMO at Millimeter Wave,'' \emph{IEEE Trans. Green Commun. Netw.}, vol. 3, no. 3, pp. 651-663, Sept. 2019.

\bibitem{introduction19}
X. Ma, Z. Chen, W. Chen \emph{et al.}, ``Joint Channel Estimation and Data Rate Maximization for Intelligent Reflecting Surface Assisted Terahertz MIMO Communication Systems,'' \emph{IEEE Access}, vol. 8, pp. 99565-99581, May 2020.

\bibitem{System01}
K. Shen and W. Yu, ``Fractional programming for communication systems-Part I: Power control and beamforming,'' \emph{IEEE Trans. Signal Process.}, vol. 66, no. 10, pp. 2616-2630, May 2018.

\bibitem{System01a}
C. L. Byrne, ``Alternating Minimization as Sequential Unconstrained Minimization: A Survey,'' \emph{J. Optim. Theory Appl.}, vol. 156, pp. 554-566, Mar. 2013.

\bibitem{System02}
S. Boyd, N. Parikh, E. Chu et al., ``Distributed optimization and statistical learning via the alternating direction method of multipliers,'' \emph{Found. Trends Machine learning}, vol. 3, no. 1, pp. 1-122, 2011.

\bibitem{System020}
Y. Liu, W. Xu, G. Wu \emph{et al.}, ``Communication-Censored ADMM for Decentralized Consensus Optimization,'' \emph{IEEE Trans. Signal Process.}, vol. 67, no. 10, pp. 2565-2579, May 2019.

\bibitem{System021}
Y. Ye, H. Chen, M. Xiao \emph{et al.}, ``Privacy-Preserving Incremental ADMM for Decentralized Consensus Optimization,'' \emph{IEEE Trans. Signal Process.}, vol. 68, pp. 5842-5854, 2020.

\bibitem{System02a}
X. Yu, J. Shen, J. Zhang \emph{et al.}, ``Alternating Minimization Algorithms for Hybrid Precoding in Millimeter Wave MIMO Systems,'' \emph{IEEE J. Sel. Top. Signal Process.}, vol. 10, no. 3, pp. 485-500, Apr. 2016.

\bibitem{System02b}
B. Nicolas, An introduction to optimization on smooth manifolds. \emph{Cambridge University Press}, Apr. 2022.

%

\bibitem{System03}
S. Boyd, ``Subgradient methods,'' pp. 1-39, May 2014. [Online]. Available: \url{https://web.stanford.edu/class/ee364b/lectures/subgrad_method_notes.pdf}


\bibitem{PCF01}
L. J. Watters, ``Reduction of integer polynomial programming problems to zero-one linear programming problems,'' \emph{Operations Research}, vol. 15, pp. 1171-1174, 1967.

\bibitem{PCF02}
L. Wolsey, Integer Programming. \emph{John Wiley and Sons}, 1998.

\bibitem{PCF03}
A. Basu, M. Conforti, M. Di Summa \emph{et al.}, ``Complexity of branch-and-bound and cutting planes in mixed-integer optimization,'' \emph{Math. Program.}, pp. 1-24, Mar. 2022.

\bibitem{Simulation01}
M. R. Akdeniz, Y. Liu, M. K. Samimi \emph{et al.}, ``Millimeter Wave Channel Modeling and Cellular Capacity Evaluation,'' \emph{IEEE J. Sel. Areas Commun.}, vol. 32, no. 6, pp. 1164-1179, June 2014.


%
%
%
%
%
%
%
%
%
%
%
%
%
%
%
%
%
%
%
%
%
%
%
%
%
%
%
%
%
%
%
%
%
%
%
%

%
%



%
%
%
%
%
%
%
%
%
%
%

\end{thebibliography}
\end{document}